\documentclass[final,5p,times,twocolumn,authoryear]{elsarticle}
\biboptions{authoryear}
\journal{Molecular Astrophysics}
\usepackage{color}
\usepackage[table]{xcolor}

\begin{document}
\begin{frontmatter}

\title{The Possibility of Forming Propargyl Alcohol in the Interstellar Medium}

\author{Prasanta Gorai$^{1}$, Ankan Das$^{1*}$, Liton Majumdar$^{2,1}$, Sandip Kumar Chakrabarti$^{3,1}$, Bhalamurugan Sivaraman$^{4}$, Eric Herbst$^{5}$}

\address{$^{1}$Indian Centre for Space Physics, Chalantika 43, Garia Station Rd., Kolkata, 700084, India.}
\address{$^{2}$Laboratoire d'astrophysique de Bordeaux, Univ. Bordeaux, CNRS, B18N, allée Geoffroy Saint-Hilaire, 33615 Pessac, France.}
\address{$^{3}$S. N. Bose National Centre for Basic Sciences, Salt Lake, Kolkata, 700098, India.}
\address{$^{4}$ Atomic Molecular and Optical Physics Division, Physical Research Laboratory, Ahmedabad, 380009, India.}
\address{$^{5}$Departments of Chemistry and Astronomy, University of Virginia, Charlottesville, VA 22904, USA.}

\cortext[$^{$*$}$]{Correspondence author: Ankan Das; Electronic mail:ankan.das@gmail.com}

\begin{abstract}
Propargyl alcohol ($\mathrm{HC_2CH_2OH}$, PA) has yet to be observed in the interstellar medium (ISM) although
one of its stable isomers, propenal ($\mathrm{CH_2CHCHO}$),  has already been detected in Sagittarius B2(N)
with the 100-meter Green Bank Telescope in the frequency range $18-26$ GHz. 
In this paper, we investigate the  formation of  propargyl alcohol along with one of its
deuterated isotopomers, $\mathrm{HC_2CH_2OD}$ (OD-PA), in a dense molecular cloud. 
Various pathways for the formation of 
PA in the gas and on ice mantles surrounding dust particles are discussed.  
We use a large gas-grain chemical network to 
study the chemical evolution of PA and its deuterated isotopomer.  
Our results suggest that  gaseous $\mathrm{HC_2CH_2OH}$ 
can most likely be detected  in hot cores or in collections 
of hot cores such as the star-forming region Sgr B2(N). 
A simple LTE (Local thermodynamic equilibrium) radiative transfer model is
employed to check the possibility of detecting PA and OD-PA in the millimeter-wave regime. 
In addition, we have carried out
quantum chemical calculations to compute the vibrational transition frequencies and intensities of these species in the infrared for perhaps future use in studies with the
James Webb Space Telescope (JWST). 
\end{abstract}


\end{frontmatter}

\section{Introduction}
\label{sec:intro}
The discovery of large numbers of  interstellar and circumstellar species 
regularly refreshes our understanding of the physical conditions of the sources of 
astrochemical interest \citep{herb06}.  Astronomical observations along with  laboratory investigations 
of various meteoritic samples have discovered the presence of numerous organic molecules of 
biological interest \citep{cron93}. It is also believed that the production of such molecules in star-
and planet-forming regions of interstellar clouds, which tend to be partially saturated species 
containing the elements nitrogen and/or oxygen in addition to carbon and hydrogen, should be connected in some manner with the production
of terrestrial bio-molecules. Other types of organic molecules are also present in the ISM. 
There is strong evidence for species astronomers refer to as ``carbon chains" in cold and dense
interstellar clouds.  These carbon chains are unsaturated and linear species, which can be 
simple hydrocarbons or species with
other heavy atoms such as cyanopolyynes, which contain a terminal cyano (CN) group.     Various 
infrared (UIR) emission bands in the $3-15$ $\mu m$ range have been observed
in different astrophysical sources \citep{alla85,tiel87}. Laboratory investigations along
with  theoretical calculations led to the hypothesis that the carriers of these bands are aromatic in nature, consisting most probably  
of free molecular polycyclic aromatic hydrocarbons (PAHs), possibly with other heavy atoms 
such as nitrogen 
\citep{sala14,nobl15}. 
Other suggestions include surface functional groups on small grains, 
 quenched carbonaceous composites, 
amorphous carbon, hydrogenated amorphous carbon and condensed phase PAHs 
\citep{bren92, jage09}.  
Recently, two fullerenes,
C$_{60}$ and C$_{70}$,  have been discovered in infrared emission in post-stellar objects \citep{cami10} while the cation C$_{60}^{+}$ has been confirmed in near-infrared absorption in a diffuse cloud \citep{walk15}.

In order to understand the synthesis of PAHs, either in interstellar or circumstellar regions,
it is essential to understand 
the formation of the six-member aromatic species, benzene 
($\mathrm{C_6H_6}$). So far there are only two experimentally studied pathways that might result in the synthesis
of interstellar or circumstellar benzene.  The first is the addition of three 
acetylene ($\mathrm{C_2H_2}$) molecules \citep{zhou10} and the second is the 
recombination of two propargyl (C$_{3}$H$_{3}$) radicals \citep{wils03}.  The formation of these radicals could occur in a number of ways.
 \cite{shar14}  carried out an experiment to study the 
thermal decomposition of PA, and found the products to include  
 $\mathrm{OH}$ and $\mathrm{C_3H_3}$, suggesting that PA could be a precursor to benzene formation.
In addition, \cite{siva15} found that benzene is the major product from PA 
irradiation, and suggested that the dissociation of PA plays a key role in the 
synthesize of benzene in  interstellar icy mantles.

Since PA might play a crucial role in the formation of PAH molecules, it is of interest
to explore various aspects of its interstellar chemistry and spectroscopy  
in detail. Although PA has not been detected unambiguously in the ISM, 
propenal (CH$_{2}$CHCHO), one of it isomers,
has  been detected \citep{holl04} towards the star-forming region Sgr B2(N). 
\cite{requ08} estimated the abundance of CH$_{2}$CHCHO to be around $0.3-2.3 \times 10^{-9}$ with respect to the
$\rm {H_2}$ molecule in the galactic center.
Moreover, PA has a well-known rotational spectrum.  
Depending upon the internal motion of the OH group, PA could possess two 
stable conformers,  named $gauche$ and $trans$. 
However, microwave studies of PA show that the molecule exists only
as the $gauche$ isomer, in which the hydroxyl H atom lies $\sim 60^{\circ}$ 
out of the $\mathrm{H-C\equiv C-C-O}$ plane \citep{hiro68}.  \cite{pear05} summarized other rotational studies of PA, extended the experimental
work of \cite{hiro68} through 600 GHz  and  obtained rotational and distortional constants for the gauche form of PA and its -OD singly deuterated 
isotopomer.    According to their studies, the $gauche$ state is split by inversion into two states, separated by 652.4 GHz for normal PA and 213.5 GHz for the -OD isotopomer.  Other spectroscopic work on PA and related species has also been undertaken. \cite{nyqu71} recorded and assigned Infrared and 
Raman spectra for PA, and its deuterated isotopomers, while 
 \cite{deve13} carried out experiments to 
determine the structure of the Ar..PA complex and its two deuterated isotopologues.
They found PA to have a $gauche$ structure, with Ar  located in between the  -OH and -C$\equiv$C-H groups.
In another study,  \cite{deve14} carried out experiments for the pure rotational spectra of the PA 
dimer and its three deuterium isotopologues.

\begin{table*} 
\vbox{
\vskip -3cm
\addtolength{\tabcolsep}{-5pt}
\centering{
\scriptsize
\caption{Formation and destruction pathways of PA and its related species.}
\renewcommand{\arraystretch}{0}
\begin{tabular}{|p{0.8in}|p{2.6in}|p{0.7in} p{0.35in} p{0.3in} p{1.1in}|p{1.2in}|}
\hline
{\bf Reaction number}&{\bf Reaction } &{\leavevmode\color{black}$\alpha$}&{\leavevmode\color{black}$\beta$}&{\leavevmode\color{black}$\gamma$}&{\bf Gas phase}& {\bf Ice phase} \\
{\bf (type)}&&&&&{\bf rate coefficient }& {\bf rate coefficient }\\
{}&&&&&{\bf at $T=10$ K ($100$ K)}& {\bf at $T=10$ K ($100$K)}\\
&&&&&&\\
\multicolumn{3}{p{2in}}{\bf Formation pathways}\\
&&&&&&\\
R1 (NR)&$\mathrm{O(^{3}P)+C_3H_3\rightarrow HC_{2}CHO+H}$($-252.30^a$,$-231.11^b$)  &{\leavevmode\color{black}$2.3 \times 10^{-10}$} &{\leavevmode\color{black}0.0}&{\leavevmode\color{black}0.0}&$2.3 \times 10^{-10}$($2.3 \times 10^{-10}$)$^x$&$2.83\times 10^{-7}$($2.51 \times 10^{4}$)$^y$\\
R2 (NR)&$\mathrm{C_2H+H_2CO\rightarrow HC_{2}CHO+H}$($-109.17^a$,$-106.307^b$) &{\leavevmode\color{black}$1.00 \times 10^{-10}$} &{\leavevmode\color{black}0.0}&{\leavevmode\leavevmode\color{black}0.0}& $1.00 \times 10^{-10}$($1.00 \times 10^{-10}$)$^x$&$2.84\times 10^{-17}$($3.03\times 10^{3}$)$^y$\\
R3 (NR)&$\mathrm{HC_2CHO+H\rightarrow HC_2CH_2O}$  ($-86.69^a$,$-86.23^b$)&{\leavevmode\color{black}-}&{\leavevmode\color{black}-}&{\leavevmode\color{black}-}& $0$($1.96 \times 10^{-15}$)$^x$&$6.24 \times 10^{-9}$  ($8.94\times 10^{-3}$)$^y$\\
R4 (RR)&$\mathrm{HC_2CH_2O+H\rightarrow HC_{2}CH_2OH}$ ($-420.35^a$,$-421.67^b$) &{\leavevmode\color{black}$1.00 \times 10^{-10}$} &{\leavevmode\color{black}0.0}&{\leavevmode\color{black}0.0}&$1.00 \times 10^{-10}$($1.00 \times 10^{-10}$)$^x$&$1.77 \times 10^{-1}$ ($2.54\times 10^{5}$)$^y$\\
R5 (RR)&$\mathrm{C_3H_3+OH\rightarrow HC_2CH_2OH}$ ($-301.35^a$,$-298.97^b$)&{\leavevmode\color{black}$1.00 \times 10^{-10}$}&{\leavevmode\color{black}0.0}&{\leavevmode\color{black}0.0}&   $1.00\times 10^{-10}$($1.00 \times 10^{-10}$)$^x$&$7.86 \times 10^{-29}$($2.19\times 10^{2}$)$^y$ \\
{ R6 (DR)}&$\mathrm{ {HC_2CH_2OH_2}^+ + e^- \rightarrow HC_2CH_2OH+H}$  &{\leavevmode\color{black}$2.67 \times 10^{-8}$}&{\leavevmode\color{black}$-0.59$}&{\leavevmode\color{black}0.0}& { $2.00 \times 10^{-7}$($5.16 \times 10^{-8}$)$^x$} &-\\
&&&&&&\\
\multicolumn{3}{p{2in}}{\bf Destruction pathways}\\
&&&&&&\\
{ R7 (NR)}&$\mathrm{ {HC_2CH_2OH} + OH \rightarrow  HOC_2HHCHOH}$  &{\leavevmode\color{black}$9.20 \times 10^{-12}$}&{\leavevmode\color{black}0.0}&{\leavevmode\color{black}0.0}& $9.20 \times 10^{-12}$($9.20 \times 10^{-12}$)$^x$&$3.58 \times 10^{-38}$($3.64\times 10^{1}$)$^y$  \\
{ R8 (NR)}&$\mathrm{ {HC_2CH_2OH} + OD \rightarrow  HOC_2HHCHOD}$  &{\leavevmode\color{black}$9.20 \times 10^{-12}$}&{\leavevmode\color{black}0.0}&{\leavevmode\color{black}0.0}& $9.20 \times 10^{-12}$($9.20 \times 10^{-12}$)$^x$&$3.48 \times 10^{-38}$($3.54\times 10^{1}$)$^y$ \\
R9 (IN)&$\mathrm{HC_2CH_2OH + C^{+} \rightarrow HC_2CH_2O^{+} + CH}$   &{\leavevmode\color{black}$2.03 \times 10^{-09}$}&{\leavevmode\color{black}-0.5}&{\leavevmode\color{black}0.0}&  $1.22\times 10^{-08}$($ 4.63 \times 10^{-09}$)$^x$&-\\
R10 (IN)&$\mathrm{HC_2CH_2OH + C^{+} \rightarrow C_3H_3^{+} + HCO}$  &{\leavevmode\color{black}$2.03 \times 10^{-09}$}&{\leavevmode\color{black}-0.5}&{\leavevmode\color{black}0.0}&  $1.22\times 10^{-08}$($ 4.61\times 10^{-09}$)$^x$&-\\
R11 (IN)&$\mathrm{HC_2CH_2OH + H_3O^{+}\rightarrow {HC_2CH_2OH_2}^{+} + H_2O}$  &{\leavevmode\color{black}$1.70 \times 10^{-09}$}&{\leavevmode\color{black}-0.5}&{\leavevmode\color{black}0.0}& $1.02\times 10^{-08}$($ 3.85\times 10^{-09}$)$^x$&- \\
R12 (IN)&$\mathrm{HC_2CH_2OH + HCO^{+}\rightarrow {HC_2CH_2OH_2}^{+} + CO }$  &{\leavevmode\color{black}$1.46 \times 10^{-09}$}&{\leavevmode\color{black}-0.5}&{\leavevmode\color{black}0.0}&  $8.79\times 10^{-09}$($ 3.32 \times 10^{-09}$)$^x$&- \\
R13 (IN)&$\mathrm{HC_2CH_2OH + H_3^{+}\rightarrow C_3H_3^{+} + H_2O + H_2}$   &{\leavevmode\color{black}$3.78 \times 10^{-09}$}&{\leavevmode\color{black}-0.5}&{\leavevmode\color{black}0.0}& $2.27\times 10^{-08}$($ 8.56\times 10^{-09}$)$^x$&- \\
R14 (IN)&$\mathrm{HC_2CH_2OH + H_3^{+}\rightarrow {HC_2CH_2OH_2}^{+} + H_2}$   &{\leavevmode\color{black}$3.78 \times 10^{-09}$}&{\leavevmode\color{black}-0.5}&{\leavevmode\color{black}0.0}& $2.27\times 10^{-08}$($ 8.56\times 10^{-09}$)$^x$&- \\
R15 (IN)&$\mathrm{HC_2CH_2OH + He^{+}\rightarrow C_3H_3^{+} + He + OH}$  &{\leavevmode\color{black}$3.31 \times 10^{-09}$}&{\leavevmode\color{black}-0.5}&{\leavevmode\color{black}0.0}& $ 1.99\times 10^{-08}$($7.51\times 10^{-09}$)$^x$&- \\
R16 (IN)&$\mathrm{HC_2CH_2OH + He^{+}\rightarrow C_3H_3 + He + OH^+}$   &{\leavevmode\color{black}$3.31 \times 10^{-09}$}&{\leavevmode\color{black}-0.5}&{\leavevmode\color{black}0.0}& $ 1.99\times 10^{-08}$($ 7.51 \times 10^{-09}$)$^x$&- \\
R17 (IN)&$\mathrm{HC_2CH_2OH + CH_3^{+}\rightarrow HC_2CH_2O^{+} + CH_4}$  &{\leavevmode\color{black}$1.86 \times 10^{-09}$}&{\leavevmode\color{black}-0.5}&{\leavevmode\color{black}0.0}&  $1.12\times 10^{-08}$($ 4.21\times 10^{-09}$)$^x$&- \\
R18 (IN)&$\mathrm{HC_2CH_2OH + H^{+}\rightarrow C_3H_3^{+} + H_2O}$   &{\leavevmode\color{black}$6.43 \times 10^{-09}$}&{\leavevmode\color{black}-0.5}&{\leavevmode\color{black}0.0}& $3.86\times 10^{-08}$($ 1.46 \times 10^{-08}$)$^x$&- \\
R19 (IN)&$\mathrm{HC_2CH_2OH + H^{+}\rightarrow HC_2CH_2O^{+} + H_2}$   &{\leavevmode\color{black}$6.43 \times 10^{-09}$}&{\leavevmode\color{black}-0.5}&{\leavevmode\color{black}0.0}& $3.86\times 10^{-08}$($ 1.46\times 10^{-08}$)$^x$&- \\
R20 (IN)&$\mathrm{HC_2CH_2OH + H^{+}\rightarrow HC_3O^{+} + H_2+H_2}$  &{\leavevmode\color{black}$6.43 \times 10^{-09}$}&{\leavevmode\color{black}-0.5}&{\leavevmode\color{black}0.0}& $3.86\times 10^{-08}$($ 1.46 \times 10^{-08}$)$^x$&- \\
R21 (IN)&$\mathrm{HC_2CH_2OH + H^{+}\rightarrow HC_2CH_2OH^{+} + H}$   &{\leavevmode\color{black}$6.43 \times 10^{-09}$}&{\leavevmode\color{black}-0.5}&{\leavevmode\color{black}0.0}& $3.86\times 10^{-08}$($ 1.46\times 10^{-08}$)$^x$&- \\
R22 (IN)&$\mathrm{HC_2CH_2OH + H_2D^{+}\rightarrow HC_2CH_2OHD^+ + H_2}$   &{\leavevmode\color{black}$3.21 \times 10^{-09}$}&{\leavevmode\color{black}-0.5}&{\leavevmode\color{black}0.0}& $1.93\times 10^{-08}$($ 7.29\times 10^{-09}$)$^x$&- \\
R23 (IN)&$\mathrm{HC_2CHO + C^{+} \rightarrow HC_3O^{+} + CH}$  &{\leavevmode\color{black}$3.80 \times 10^{-09}$}&{\leavevmode\color{black}-0.5}&{\leavevmode\color{black}0.0}&  $2.19\times 10^{-08}$($7.64 \times 10^{-09}$)$^x$&-\\
R24 (IN)&$\mathrm{HC_2CHO + C^{+} \rightarrow C_3H^{+} + HCO}$  &{\leavevmode\color{black}$3.80 \times 10^{-09}$}&{\leavevmode\color{black}-0.5}&{\leavevmode\color{black}0.0}&  $2.19\times 10^{-08}$($ 7.64 \times 10^{-09}$)$^x$&-\\
R25 (IN)&$\mathrm{HC_2CHO + H_3O^{+}\rightarrow {HC_2CH_2O}^{+} + H_2O}$   &{\leavevmode\color{black}$3.17 \times 10^{-09}$}&{\leavevmode\color{black}-0.5}&{\leavevmode\color{black}0.0}& $1.83\times 10^{-08}$($ 6.38 \times 10^{-09}$)$^x$&- \\
R26 (IN)&$\mathrm{HC_2CHO + HCO^{+}\rightarrow {HC_2CH_2O}^{+} + CO }$  &{\leavevmode\color{black}$2.74 \times 10^{-09}$}&{\leavevmode\color{black}-0.5}&{\leavevmode\color{black}0.0}&  $1.58\times 10^{-08}$($ 5.51\times 10^{-09}$)$^x$&- \\
R27 (IN)&$\mathrm{HC_2CHO + H_3^{+}\rightarrow C_3H^{+} + H_2O + H_2}$   &{\leavevmode\color{black}$7.03 \times 10^{-09}$}&{\leavevmode\color{black}-0.5}&{\leavevmode\color{black}0.0}& $4.04\times 10^{-08}$($ 1.41 \times 10^{-08}$)$^x$&- \\
R28 (IN)&$\mathrm{HC_2CHO + H_3^{+}\rightarrow {HC_2CH_2O}^{+} + H_2}$   &{\leavevmode\color{black}$7.03 \times 10^{-09}$}&{\leavevmode\color{black}-0.5}&{\leavevmode\color{black}0.0}& $4.04\times 10^{-08}$($ 1.41\times 10^{-08}$)$^x$&- \\
R29 (IN)&$\mathrm{HC_2CHO + He^{+}\rightarrow C_3H^{+} + He + OH}$  &{\leavevmode\color{black}$6.17 \times 10^{-09}$}&{\leavevmode\color{black}-0.5}&{\leavevmode\color{black}0.0}& $ 3.55\times 10^{-08}$($ 1.24\times 10^{-08}$)$^x$&- \\
R30 (IN)&$\mathrm{HC_2CHO + He^{+}\rightarrow C_3H + He + OH^+}$   &{\leavevmode\color{black}$6.17 \times 10^{-09}$}&{\leavevmode\color{black}-0.5}&{\leavevmode\color{black}0.0}& $ 3.55\times 10^{-08}$($ 1.24 \times 10^{-08}$)$^x$&- \\
R31 (IN)&$\mathrm{HC_2CHO + CH_3^{+}\rightarrow HC_2CH_2O^{+} + CH_2}$  &{\leavevmode\color{black}$3.47 \times 10^{-09}$}&{\leavevmode\color{black}-0.5}&{\leavevmode\color{black}0.0}&  $2.00\times 10^{-08}$ ( $ 6.98 \times 10^{-09}$)$^x$&- \\
R32 (IN)&$\mathrm{HC_2CHO + H^{+}\rightarrow C_3H^{+} + H_2O}$   &{\leavevmode\color{black}$1.20 \times 10^{-08}$}&{\leavevmode\color{black}-0.5}&{\leavevmode\color{black}0.0}& $6.89\times 10^{-08}$ ( $ 2.41 \times 10^{-08}$)$^x$&- \\
R33 (IN)&$\mathrm{HC_2CHO + H^{+}\rightarrow C_3H_2 + OH^+}$   &{\leavevmode\color{black}$1.20 \times 10^{-08}$}&{\leavevmode\color{black}-0.5}&{\leavevmode\color{black}0.0}& $6.89\times 10^{-08}$ ( $ 2.41 \times 10^{-08}$)$^x$&- \\
R34 (IN)&$\mathrm{HC_2CHO + H^{+}\rightarrow HC_3O^{+} + H_2}$  &{\leavevmode\color{black}$1.20 \times 10^{-08}$}&{\leavevmode\color{black}-0.5}&{\leavevmode\color{black}0.0}& $6.89\times 10^{-08}$ ( $ 2.41 \times 10^{-08}$)$^x$&- \\
R35 (IN)&$\mathrm{HC_2CHO + H^{+}\rightarrow C_3O^{+} + H_2}$   &{\leavevmode\color{black}$1.20 \times 10^{-08}$}&{\leavevmode\color{black}-0.5}&{\leavevmode\color{black}0.0}& $6.89\times 10^{-08}$ ($ 2.41\times 10^{-08}$)$^x$&- \\
R36 (IN)&$\mathrm{HC_2CHO + H_2D^{+}\rightarrow HC_2CH_2O^+ + HD}$  &{\leavevmode\color{black}$5.98 \times 10^{-09}$}&{\leavevmode\color{black}-0.5}&{\leavevmode\color{black}0.0}& $3.44\times 10^{-08}$ ($ 1.20\times 10^{-08}$)$^x$&- \\
R37 (IN)&$\mathrm{HC_2CH_2O + C^{+} \rightarrow HC_3O^{+} + CH_2}$   &{\leavevmode\color{black}$1.41 \times 10^{-09}$}&{\leavevmode\color{black}-0.5}&{\leavevmode\color{black}0.0}&  $8.86\times 10^{-09}$ ($ 3.57 \times 10^{-09}$)$^x$&-\\
R38 (IN)&$\mathrm{HC_2CH_2O + C^{+} \rightarrow C_3H_2^{+} + HCO}$  &{\leavevmode\color{black}$1.41 \times 10^{-09}$}&{\leavevmode\color{black}-0.5}&{\leavevmode\color{black}0.0}&  $8.86\times 10^{-09}$ ($ 3.57 \times 10^{-09}$)$^x$&-\\
R39 (IN)&$\mathrm{HC_2CH_2O + H_3O^{+}\rightarrow {HC_2CH_2O}^{+} + H_2O}$   &{\leavevmode\color{black}$1.18 \times 10^{-09}$}&{\leavevmode\color{black}-0.5}&{\leavevmode\color{black}0.0}& $7.40\times 10^{-09}$ ($ 2.92 \times 10^{-09}$)$^x$&- \\
R40 (IN)&$\mathrm{HC_2CH_2O + HCO^{+}\rightarrow {HC_2CH_2O}^{+} + HCO }$  &{\leavevmode\color{black}$1.02 \times 10^{-09}$}&{\leavevmode\color{black}-0.5}&{\leavevmode\color{black}0.0}&  $6.38\times 10^{-09}$ ($ 2.58 \times 10^{-09}$)$^x$&- \\
R41 (IN)&$\mathrm{HC_2CH_2O + H_3^{+}\rightarrow C_3H_2^{+} + H_2O + H_2}$   &{\leavevmode\color{black}$2.62 \times 10^{-09}$}&{\leavevmode\color{black}-0.5}&{\leavevmode\color{black}0.0}& $1.64\times 10^{-08}$ ($ 6.62 \times 10^{-09}$)$^x$&- \\
R42 (IN)&$\mathrm{HC_2CH_2O + H_3^{+}\rightarrow {HC_2CH_2O}^{+} + H_2+H}$   &{\leavevmode\color{black}$2.62 \times 10^{-09}$}&{\leavevmode\color{black}-0.5}&{\leavevmode\color{black}0.0}& $1.64\times 10^{-08}$ ($ 6.62 \times 10^{-09}$)$^x$&- \\
R43 (IN)&$\mathrm{HC_2CH_2O + He^{+}\rightarrow C_3H^{+} + He + H_2O}$   &{\leavevmode\color{black}$2.30 \times 10^{-09}$}&{\leavevmode\color{black}-0.5}&{\leavevmode\color{black}0.0}& $ 1.44\times 10^{-08}$ ($ 5.80 \times 10^{-09}$)$^x$&- \\
R44 (IN)&$\mathrm{HC_2CH_2O + He^{+}\rightarrow C_4H + He + OH^+}$   &{\leavevmode\color{black}$2.30 \times 10^{-09}$}&{\leavevmode\color{black}-0.5}&{\leavevmode\color{black}0.0}& $ 1.44\times 10^{-08}$ ($ 5.80 \times 10^{-09}$)$^x$&- \\
R45 (IN)&$\mathrm{HC_2CH_2O + CH_3^{+}\rightarrow HC_2CH_2O^{+} + CH_3}$  &{\leavevmode\color{black}$1.29 \times 10^{-09}$}&{\leavevmode\color{black}-0.5}&{\leavevmode\color{black}0.0}&  $8.09\times 10^{-09}$ ($ 3.26 \times 10^{-09}$)$^x$&- \\
R46 (IN)&$\mathrm{HC_2CH_2O + H^{+}\rightarrow C_3H_2^{+} + H_2O}$   &{\leavevmode\color{black}$4.46 \times 10^{-09}$}&{\leavevmode\color{black}-0.5}&{\leavevmode\color{black}0.0}& $2.80\times 10^{-08}$ ($ 1.13 \times 10^{-08}$)$^x$&- \\
R47 (IN)&$\mathrm{HC_2CH_2O + H^{+}\rightarrow C_3H_2 +OH^+ + H}$   &{\leavevmode\color{black}$4.46 \times 10^{-09}$}&{\leavevmode\color{black}-0.5}&{\leavevmode\color{black}0.0}& $2.80\times 10^{-08}$ ($ 1.13 \times 10^{-08}$)$^x$&- \\
R48 (IN)&$\mathrm{HC_2CH_2O + H^{+}\rightarrow HC_3O^{+} + H_2+H}$  &{\leavevmode\color{black}$4.46 \times 10^{-09}$}&{\leavevmode\color{black}-0.5}&{\leavevmode\color{black}0.0}& $2.80\times 10^{-08}$ ($ 1.13 \times 10^{-08}$)$^x$&- \\
R49 (IN)&$\mathrm{HC_2CH_2O + H^{+}\rightarrow HC_3O^{+} + H_2}$   &{\leavevmode\color{black}$4.46 \times 10^{-09}$}&{\leavevmode\color{black}-0.5}&{\leavevmode\color{black}0.0}& $2.80\times 10^{-08}$ ($1.13\times 10^{-08}$)$^x$&- \\
R50 (IN)&$\mathrm{HC_2CH_2O + H_2D^{+}\rightarrow HC_2CH_2O^+ + HD + H}$   &{\leavevmode\color{black}$2.23 \times 10^{-09}$}&{\leavevmode\color{black}-0.5}&{\leavevmode\color{black}0.0}& $1.40\times 10^{-08}$ ($ 5.63\times 10^{-09}$)$^x$&- \\
R51 (PH)&$\mathrm{HC_2CH_2OH + PHOTON \rightarrow C_3H_3 + OH}$ &{\leavevmode\color{black}$6.0\times 10^{-10}$}&{\leavevmode\color{black}0.0}&{\leavevmode\color{black}1.8}& $9.14 \times 10^{-18}$($9.14 \times 10^{-18}$)$^y$&$9.14\times 10^{-18}$ ($9.14\times 10^{-18}$)$^y$\\
R52 (PH)&$\mathrm{ HC_2CH_2OH + PHOTON \rightarrow HC_2CH_2O + H}$ &{\leavevmode\color{black}$6.0\times 10^{-10}$}&{\leavevmode\color{black}0.0}&{\leavevmode\color{black}1.8}&  $9.14 \times 10^{-18}$($9.14 \times 10^{-18}$)$^y$&$9.14\times 10^{-18}$ ($9.14\times 10^{-18}$)$^y$\\
R53 (PH)&$\mathrm{HC_2CH_2O + PHOTON \rightarrow C_3H_2 + OH}$ &{\leavevmode\color{black}$6.0\times 10^{-10}$}&{\leavevmode\color{black}0.0}&{\leavevmode\color{black}1.8}&$9.14 \times 10^{-18}$($9.14 \times 10^{-18}$)$^y$&$9.14\times 10^{-18}$ ($9.14\times 10^{-18}$)$^y$\\
R54 (PH)&$\mathrm{HC_2CH_2O + PHOTON \rightarrow H_2CO + C_2H }$ &{\leavevmode\color{black}$6.0\times 10^{-10}$}&{\leavevmode\color{black}0.0}&{\leavevmode\color{black}1.8}& $9.14 \times 10^{-18}$($9.14 \times 10^{-18}$)$^y$&$9.14\times 10^{-18}$ ($9.14\times 10^{-18}$)$^y$\\
R55 (PH)&$\mathrm{ HC_2CH_2O + PHOTON \rightarrow HC_2CHO + H }$ &{\leavevmode\color{black}$6.0\times 10^{-10}$ }&{\leavevmode\color{black}0.0}&{\leavevmode\color{black}1.8}& $9.14 \times 10^{-18}$($9.14 \times 10^{-18}$)$^y$&$9.14\times 10^{-18}$ ($9.14\times 10^{-18}$)$^y$\\
R56 (PH)&$\mathrm{HC_2CHO + PHOTON \rightarrow C_2H + HCO }$ &{\leavevmode\color{black} $6.0\times 10^{-10}$}&{\leavevmode\color{black}0.0}&{\leavevmode\color{black}1.8}& $9.14 \times 10^{-18}$($9.14 \times 10^{-18}$)$^y$&$9.14\times 10^{-18}$ ($9.14\times 10^{-18}$)$^y$\\
R57 (PH)&$\mathrm{HOC_2HHCHOH + PHOTON \rightarrow HC_2CH_2OH + OH}$ &{\leavevmode\color{black}$6.0\times 10^{-10}$}&{\leavevmode\color{black}0.0}&{\leavevmode\color{black}1.8}&  $9.14 \times 10^{-18}$($9.14 \times 10^{-18}$)$^y$&$9.14\times 10^{-18}$ ($9.14\times 10^{-18}$)$^y$\\
R58 (PH)&$\mathrm{HOC_2HHCHOD + PHOTON \rightarrow HC_2CH_2OH + OD}$ &{\leavevmode\color{black}$6.0\times 10^{-10}$}&{\leavevmode\color{black}0.0}&{\leavevmode\color{black}1.8}&  $9.14 \times 10^{-18}$($9.14 \times 10^{-18}$)$^y$&$9.14\times 10^{-18}$ ($9.14\times 10^{-18}$)$^y$\\
R59 (CR)&$\mathrm{  HC_2CH_2OH + CRPHOT \rightarrow C_3H_3 + OH}$  &{\leavevmode\color{black}$1.3\times 10^{-17}$}&{\leavevmode\color{black}0.0}&{\leavevmode\color{black}752}& $2.44 \times 10^{-14}$($2.44 \times 10^{-14}$)$^y$&$2.44 \times 10^{-14}$ ($2.44\times 10^{-14}$)$^y$\\
R60 (CR)&$\mathrm{ HC_2CH_2OH + CRPHOT \rightarrow HC_2CH_2O + H}$  &{\leavevmode\color{black}$1.3\times 10^{-17}$}&{\leavevmode\color{black}0.0}&{\leavevmode\color{black}752}& $2.44 \times 10^{-14}$($2.44 \times 10^{-14}$)$^y$&$2.44 \times 10^{-14}$ ($2.44\times 10^{-14}$)$^y$\\
R61 (CR)&$\mathrm{ HC_2CH_2O + CRPHOT \rightarrow C_3H_2 + OH}$  &{\leavevmode\color{black}$1.3\times 10^{-17}$}&{\leavevmode\color{black}0.0}&{\leavevmode\color{black}752}& $2.44 \times 10^{-14}$($2.44 \times 10^{-14}$)$^y$&$2.44 \times 10^{-14}$ ($2.44\times 10^{-14}$)$^y$\\
R62 (CR)&$\mathrm{ HC_2CH_2O + CRPHOT \rightarrow H_2CO + C_2H }$ &{\leavevmode\color{black}$1.3\times 10^{-17}$}&{\leavevmode\color{black}0.0}&{\leavevmode\color{black}752}&  $2.44 \times 10^{-14}$($2.44 \times 10^{-14}$)$^y$&$2.44\times 10^{-14}$ ($2.44\times 10^{-14}$)$^y$\\
R63 (CR)&$\mathrm{ HC_2CH_2O + CRPHOT \rightarrow HC_2CHO + H }$ &{\leavevmode\color{black}$1.3\times 10^{-17}$}&{\leavevmode\color{black}0.0}&{\leavevmode\color{black}752}&  $2.44 \times 10^{-14}$($2.44 \times 10^{-14}$)$^y$&$2.44\times 10^{-14}$ ($2.44\times 10^{-14}$)$^y$\\
R64 (CR)&$\mathrm{ HC_2CHO + CRPHOT \rightarrow C_2H + HCO}$  &{\leavevmode\color{black}$1.3\times 10^{-17}$}&{\leavevmode\color{black}0.0}&{\leavevmode\color{black}752}& $2.44 \times 10^{-14}$($2.44 \times 10^{-14}$)$^y$&$2.44 \times 10^{-14}$ ($2.44\times 10^{-14}$)$^y$\\
R65 (CR)&$\mathrm{ HOC_2HHCHOH + CRPHOT \rightarrow HC_2CH_2OH + OH}$ &{\leavevmode\color{black}$1.3\times 10^{-17}$}&{\leavevmode\color{black}0.0}&{\leavevmode\color{black}752}&  $2.44 \times 10^{-14}$($2.44 \times 10^{-14}$)$^y$&$2.44\times 10^{-14}$ ($2.44\times 10^{-14}$)$^y$\\
R66 (CR)&$\mathrm{ HOC_2HHCHOD + CRPHOT \rightarrow HC_2CH_2OH + OD}$ &{\leavevmode\color{black}$1.3\times 10^{-17}$}&{\leavevmode\color{black}0.0}&{\leavevmode\color{black}752}&  $2.44 \times 10^{-14}$($2.44 \times 10^{-14}$)$^y$&$2.44\times 10^{-14}$ ($2.44\times 10^{-14}$)$^y$\\
R67 (DR)&$\mathrm{ HC_2CH_2O^+ + e^- \rightarrow CO+C_2H_2+H}$&{\leavevmode\color{black}$2.0\times 10^{-7}$}&{\leavevmode\color{black}-0.5}&{\leavevmode\color{black}0.0}&{$1.10 \times 10^{-6}$($3.46 \times 10^{-7}$) $^x$}&-\\
R68(DR)&$\mathrm{ HC_2CH_2O^+ + e^- \rightarrow HCO+C_2H+H}$ &{\leavevmode\color{black}$2.0\times 10^{-7}$}&{\leavevmode\color{black}-0.5}&{\leavevmode\color{black}0.0}&{$1.10 \times 10^{-6}$}($3.46 \times 10^{-7}$)$^x$&- \\
R69(DR)&$\mathrm{ HC_2CH_2O^+ + e^- \rightarrow H_2CO+C_2H}$  &{\leavevmode\color{black}$2.0\times 10^{-7}$}&{\leavevmode\color{black}-0.5}&{\leavevmode\color{black}0.0}&{$1.10 \times 10^{-6}$}($3.46 \times 10^{-7}$)$^x$&- \\
R70(DR)&$\mathrm{ {HC_2CH_2OH_2}^+ + e^- \rightarrow C_3H_3+H_2O}$ &{\leavevmode\color{black}$8.01\times 10^{-8}$}&{\leavevmode\color{black}-0.59}&{\leavevmode\color{black}0.0}&{$5.95 \times 10^{-7}$($1.52 \times 10^{-7}$)$^x$}&- \\
R71(DR)&$\mathrm{ {HC_2CH_2OH_2}^+ + e^- \rightarrow C_3H_3+OH+H}$ &{\leavevmode\color{black}$4.54\times 10^{-7}$}&{\leavevmode\color{black}-0.59}&{\leavevmode\color{black}0.0}& {$3.34 \times 10^{-6}$($8.6 \times 10^{-7}$)$^x$}&- \\
R72(DR)&$\mathrm{ {HC_2CH_2OH_2}^+ + e^- \rightarrow C_3H_2+H_2O+H}$ &{\leavevmode\color{black}$1.87\times 10^{-7}$}&{\leavevmode\color{black}-0.59}& {\leavevmode\color{black}0.0}&{$1.41 \times 10^{-6}$($3.63 \times 10^{-7}$)$^x$}&-\\
R73(DR)&$\mathrm{ {HC_2CH_2OH_2}^+ + e^- \rightarrow H_2CO+C_2H_2+H}$ &{\leavevmode\color{black}$8.9 \times 10^{-8}$} &{\leavevmode\color{black} -0.59}&{\leavevmode\color{black}0.0}& {$6.6 \times 10^{-7}$($1.7 \times 10^{-7}$)$^x$}&-\\ 
R74(DR)&$\mathrm{ {HC_2CH_2OH}^+ + e^- \rightarrow C_3H_3+OH}$ &{\leavevmode\color{black}$3.00\times 10^{-7}$}&{\leavevmode\color{black}-0.5} & {\leavevmode\color{black}0.0}&{$1.64 \times 10^{-6}$($5.20 \times 10^{-7}$)$^x$}&- \\
R75(DR)&$\mathrm{ {HC_2CH_2OH}^+ + e^- \rightarrow HC_2CH_2O+H}$ &{\leavevmode\color{black}$3.00\times 10^{-7}$}&{\leavevmode\color{black}-0.5}&{\leavevmode\color{black}0.0}&{$1.64 \times 10^{-6}$($5.20 \times 10^{-7}$)$^x$}&- \\
\hline
\multicolumn{7}{c}{ Notes: For the two-body reactions (R1-R2, R4-R8, R67-R75), rate coefficients are tabulated as $\alpha \ (\frac{T}{300})^{\beta} \ exp(-\frac{\gamma}{T})$. For the photo-dissociation reactions with}\\
\multicolumn{7}{c}{ external interstellar photons (R51-R58), rate coefficients are tabulated as $\alpha \ exp(-\gamma A_V)$. For  photo-dissociation by cosmic ray induced photons (R59-R66),} \\
\multicolumn{7}{c}{\color{black} rate coefficients are tabulated as $\alpha \ \frac{\gamma'}{1-\omega}$. For the rate coefficients of ion-polar neutral reactions (R9-R50), we use the relation discussed in \cite{su82}.} \\
\multicolumn{7}{c}{\color{black} Tabulated rate coefficients for these ion-neutral reactions in terms of $\alpha \ (\frac{T}{300})^{\beta} \ exp(-\frac{\gamma}{T})$ are valid for the low temperature regime only.}\\
\multicolumn{3}{c}{$^a$ gas phase enthalpy in kJ/mol}\\
\multicolumn{3}{c}{$^b$ ice phase enthalpy in kJ/mol}\\
\multicolumn{3}{c}{$^x$ rate coefficient in cm$^3$s$^{-1}$}\\
\multicolumn{3}{c}{$^y$ rate coefficient in s$^{-1}$}\\
\end{tabular}}}
\end{table*}


In this paper, we report the use of our interstellar chemical model to explore various pathways for the 
formation and destruction of PA (gauche form), and to estimate the possibility of detecting 
this molecule in a dense molecular cloud.  
Since there are some existing laboratory results for the spectrum of the  -OD deuterated form of 
PA and since some observational evidence for deuterium fractionation of large complex species exists
(see for instance $\mathrm{DCOOCH_3/HCOOCH_3}$, \cite{demy10}), we also consider the -OD isotopomer of PA. 
Various vibrational transitions of PA are computed and compared with the existing experimental results.

The remainder of this paper is organized as follows. In Section 2, we discuss various reaction pathways and 
their rate coefficients for the formation and destruction of PA and OD-PA.  In {\color{black} Section 3.1, 
modeling details are presented while in Section 3.2 we discuss modeling results..} {\color{black} LTE radiative transfer results are presented in 
{\color{black} Section 4.1}},
while computed vibrational spectra for PA and OD-PA are
discussed in {\color{black} Section 4.2}. Finally, in {\color{black} Section 5}, we draw our conclusions.

\section{Chemical network}
\label{sec:chem}

\subsection{Formation pathways}
In Table 1, all formation and destruction pathways of PA utilized
are presented with rate coefficients, if applicable, in both the gas and dust phases.  The rate coefficients are shown for two temperatures ($T=10$ K and $100$ K)
to represent the temperature dependency (if any). The determination of the rate coefficients actually used is discussed in the next few subsections.
Most  rate coefficients for the case of deuterated PA are not very different, and are not tabulated. Reaction numbers R1-R6 of Table 1 represent various possible 
pathways for the formation of PA ($\mathrm{HC_2CH_2OH}$). Reaction numbers R1-R5 are found to be 
exothermic in both phases and are included in our network. The reaction
exothermicities or endothermicities  for 
all these reactions have been calculated by using the Gaussian 09 \citep{fris09} program with a 
B3LYP functional  \citep{beck88,lee88} and basis set 6-311g++(d,p). 
Note that reaction exothermicities or endothermicities do not differ significantly between the 
gaseous and ice mantle phases.  We calculated the endothermicity/exothermicity ($\Delta H$) of a reaction
by taking the difference between the total optimized enthalpy including zero point corrections
of the products 
and reactants. 
If $\Delta H$ is positive, we label the reaction endothermic and if $\Delta H$ is negative, we label the reaction
exothermic. 
Another formation reaction, R6, is considered only in the gas phase.   
The individual formation reactions are discussed in the following paragraphs. 

Reaction R1 ($\mathrm{O + C_3H_3}$) in the gas phase was 
studied by \cite{kwon06}, who carried out an experiment as 
well as ab initio statistical calculations. 
They found that the reaction is barrier-less and can produce  
propynal (HC$_{2}$CHO) and H.  The conversion of propynal into PA then can occur 
via two association reactions (R3, R4) with atomic hydrogen. In the gas phase, 
the process occurs via radiative association, in which emission of a photon stabilizes the 
intermediate reaction complex. \cite{lee06} predicted that the barrier-less 
addition of O($^3$P) to propargyl radical ($\mathrm{C_3H_3}$) on the lowest doublet potential energy 
surface could produce several energy-rich intermediates, which undergo subsequent 
isomerization and decomposition steps to generate various exothermic reaction products. 
Their statistical calculation also suggests that the primary reaction channel leads to
the formation of propynal. 
Reaction R2, in which the radical CCH and formaldehyde produce propynal + H,  was studied by \cite{dong05} and by \cite{petr95}.   \cite{dong05} calculated a very small barrier of 2.1 kcal/mol at the highest level of theory, while \cite{petr95} assumed the channel to be barrier-less based on similar reactions.    The propynal product can then also  be hydrogenated to PA via R3 and R4.

As an alternative to two successive H-atom association reactions involving 
atomic hydrogen, we checked the reaction  of $\mathrm{H_2}$ with propynal ($\mathrm{HC_2CHO}$) 
to form PA but found it to be highly endothermic.  
We tried a few other pathways for the formation of PA
via single step reactions, sometimes involving a radical.
In this effort, we considered  the reaction  between $\mathrm{C_2H_4}$ and CO, the 
reaction between propynal and $\mathrm{H_2O}$,  and the reaction between 
the propargyl radical and OH (reaction number R5). The reactions between $\mathrm{C_2H_4}$ and CO and
 between propynal and $\mathrm{H_2O}$ are found to be highly endothermic 
in nature whereas reaction R5 is highly exothermic and is likely barrier-less, 
since the reactants are both radicals.   
Since both reactants in R5 ($\mathrm{C_3H_3}$ and OH) are reasonably abundant in the ISM, 
we think that this reaction can contribute towards the formation of interstellar PA, 
although, in the gas phase, it must proceed via a radiative route, so that it must be looked at closely.  
On ice mantles, however, radical-radical association reactions are normally quite efficient.  
As can be seen in Table 1, R4 and R5 are the sole radical-radical reactions leading directly to the formation of PA, 
both in the gas and on the ice. A gas phase  dissociative recombination (DR) reaction (R6), in which protonated PA and 
an electron recombine to form smaller neutral products, is also included.   This reaction  may contribute significantly to the gas phase formation
of PA unless the two-body product channel shown is unimportant if ion-neutral processes can produce protonated PA efficiently.  Similar pathways to
all those considered for the synthesis of PA are assumed to be responsible for the production of OD-PA. 

\subsection{Destruction pathways}
As shown in Table 1, the destruction of gaseous PA occurs via ion-neutral (IN) and  photo-dissociative (PH \& CR) pathways, as well as via two radical-neutral 
(NR) reactions. The destruction of gas phase PA also occurs via adsorption onto ice, but the reverse process of desorption also occurs.  These processes 
are in our network, but not listed in Table 1. The rate coefficient for the gas phase NR reaction between the hydroxyl radical (OH) and PA was measured by 
\cite{upad01}, who used laser photolysis combined with the laser induced 	fluorescence technique at room temperature.  According to their study, this 
reaction (R7) produces an adduct, $\mathrm{HOC_2HHCHOH}$. 
Since the abundances of OH and OD are comparable  in a dense cloud \citep{mill89},  a similar destruction mechanism 
with OD (R8) is also considered here. In addition to the gas-phase, reactions R7 and R8 are included in the ice phase as well. For the ion-neutral 
(IN) destruction of gas phase PA, we include reactions  R9-R22  by following the similar gas phase ion-neutral (IN) destruction pathways available 
for methanol in \cite{wood07}. Similar IN destruction reactions (R23-R50) for $\rm{HC_2CHO}$ and $\rm{HC_2CH_2O}$ are also considered.
Photo-dissociation reactions (direct or cosmic ray induced) (R51-R66) are also responsible for the destruction of PA and its associated species in both phases. 
In Table 1,  DR reactions R6 and R67-R75  involve the ions ($\mathrm{HC_2CH_2O^+}$, $\mathrm{HC_2CH_2OH_2}$$^+$ and $\mathrm{HC_2CH_2OH^+}$); these ions are 
produced by various formation/destruction pathways of PA and its related species.  For the destruction of OD-PA and other associated species similar
destruction pathways to all those considered for PA are included.

\subsection{Rate coefficients}

\subsubsection{Gas phase rate coefficients}
\cite{slag91} experimentally obtained a temperature independent rate coefficient of
$\sim 2.31 \times 10^{-10}$ cm$^3$~s$^{-1}$ 
for reaction R1. According to their study, this reaction 
proceeds through a fast and irreversible association-fragmentation process.  
We utilize the rate coefficient obtained by \cite{slag91}.
\cite{petr95} estimated a rate coefficient of $1.0 \times 10^{-10}$ 
cm$^3$~s$^{-1}$ for reaction R2 with the assumption that it occurs without a barrier in the gas phase. Although this assumption contradicts the calculation of a small barrier by \cite{dong05}, we assume the reaction to be barrier-less, and use the estimated rate coefficient of \cite{petr95}.

As can be seen in Table 1, reactions R3, R4,  and R5 are highly exothermic in nature.
Based on the high exothermicity of these reactions, one might assume that these
reactions can process without barriers.

The Hydrogen addition reaction of $\rm{HC_2CHO}$ (reaction R3) may occur in two ways. First,
$\rm{H}$ addition may occur with the $\rm{O}$ atom of the $\rm{CHO}$ group and produce
${\rm HC_2CHOH}$ and secondly H addition may occur with the $\rm{C}$ atom of the $\rm{CHO}$ group
and produce ${\rm HC_2CH_2O}$. Our quantum chemical calculation  found that the hydrogen addition 
to carbon is more favourable than the hydrogen addition to oxygen in the $\rm{CHO}$ group. 
Using the DFT/6-31+G(d,p) method, we found that the gas-phase activation barrier ($\Delta E\ddag$) and Gibbs energy
of activation ($\Delta {G}\ddag$) for the second possibility 
of reaction R3 (${\rm H+HC_2CHO\rightarrow HC_2CH_2O}$) is $3.74$ kcal/mol and $9.63$ kcal/mol respectively. 

For the computation of the gas phase rate coefficient for reaction R3, we use transition state theory,  which leads 
to the Eyring equation \citep{eyri35}:
\begin{equation}
	k= (k_BT/hc) \exp(-\Delta {G}\ddag/RT).
\end{equation}
{\leavevmode\color{black} 
The rate coefficient calculated by the above equation thus has a strong temperature dependence.  
In Table 1, we have included the rate coefficient for two temperatures; $10$ K and $100$ K. 
At $10$ K, rate coefficient is $\sim 0$ and at $100$ K it becomes $1.96 \times 10^{-15}$ cm$^3$ s$^{-1}$. 
Thus in the low temperature regime, R3 does not contribute at all to the gas phase formation of PA,
while it could play a role for the formation of PA in the high temperature regime.
}

In the case of reactions R4 and R5, we were unable to locate any suitable transition state
and assume that these two reactions are barrier-less, as is customary for radical-radical reactions. The rate coefficients of these
two reactions are assumed to be $10^{-10}$ $\rm{cm^3 \ s^{-1}}$.
The formation of OD-PA is treated with similar pathways and rate coefficients.

For the formation of PA by the DR mechanism, which involves the destruction of
$\mathrm{HC_2CH_2OH_2}^+$ (R6, R67-R75), 
we follow the destruction of
$\mathrm{CH_3OH_2}^+$ from \cite{wood07} for the  rate coefficients and product channels.  Rate coefficients between two species are standardly parameterized as a function of temperature by the equation
\begin{equation}
k = {\alpha}(T/300)^{\beta} \exp^{-\gamma/T}.
\end{equation}
In this particular case, the values of the parameters for the formation of PA are
$\alpha=2.67 \times 10^{-8}$ cm$^{3}$ s$^{-1}$,
$\beta=-0.59$ and $\gamma=0$.  Similar
product channels and rate coefficients are used for the formation of OD-PA by a DR reaction. 

Now let us consider the destruction of gaseous PA.  \cite{upad01} studied the rate coefficient of the NR reaction between PA and the OH radical (R7). According to their study,
it produces an adduct with a rate coefficient of $(9.2\pm1.4) \times 10^{-12}$
$\rm {cm^3 s^{-1}}$, which we use with no temperature dependence. Here also, we assume a similar rate coefficient
for reaction R8, which involves OD.  Similar destruction reactions and rate coefficients are adopted for
OD-PA as well.

Ion Neutral (IN) reactions are the dominant means for the destruction of
neutral interstellar species. If the neutral species is non-polar, we use the
Langevin collision rate coefficient \citep{herb06,wake10}.
If the neutral species is polar, we employ the trajectory scaling relation found in
\cite{su82} and \cite{woon09}.
From our quantum chemical calculations, we find that the polarizability $\alpha_d=5.62 \times 10^{-24}$ cm$^3$
and dipole moment $\mu_D$=1.6548 Debye for PA.
For the destruction reactions of the other two associated species (${\rm HC_2CHO,\ HC_2CH_2O}$), 
in reaction numbers R23-R36,
we use $\alpha_d=5.26 \times 10^{-24}$ cm$^3$
and $\mu_D$=3.08 Debye, and for R37-R50 we use
$\alpha_d=5.93 \times 10^{-24}$ cm$^3$
and $\mu_D$=1.1480 Debye. For deuterated PA, the ion-neutral destruction reactions have similar rate
coefficients to those of normal PA, the only differences being due to the
reduced mass, which are rather small.  Deuterated reactions and their rate coefficients
are not tabulated here.

For the photo-dissociation reactions of PA and its  associated species by external interstellar photons and cosmic ray-induced photons, we use analogous products and
the same first-order rate coefficients (s$^{-1}$)  as the case of $\mathrm{CH_3OH}$ \citep{wood07}.
For the case of external photons, 
we use the following relation for the rate coefficients:
\begin{equation}
k = \alpha \exp(-\gamma A_V )  
\end{equation}
where  $\alpha$ represents the rate coefficient (s$^{-1}$) in the unshielded interstellar ultraviolet radiation field, $A_V$ is 
the visual extinction, for which we use a value of 10,  and $\gamma$ is 
used to take into account the increased 
extinction of dust in the UV. Here, following \cite{wood07}, we use $\alpha = 6.0 \times 10^{-10}$ s$^{-1}$, 
 and $\gamma = 1.8$ in our model. 
Incorporating all the parameters into the above equation, we obtain a photo-dissociation rate coefficient of 
about $9.14 \times 10^{-18}$ s$^{-1}$.
 
For cosmic-ray-induced photo-reactions, we use the following relation, which was originally developed by \citep{gred89}:
\begin{equation}
 k=\alpha \gamma'/(1- \omega)
\end{equation}
where $\alpha$ is the cosmic-ray ionization rate (s$^{-1}$), $\gamma'$ is the number of photo-dissociative events that take place per cosmic-ray ionization 
 and $\omega$ is the dust grain albedo in the far ultraviolet. Here, we use $\omega$ =0.6,  
$\alpha=1.3 \times 10^{-17}$ s$^{-1}$, and $\gamma'=752.0$  by following the cosmic-ray-induced 
photo-reactions of $\rm{CH_3OH}$ in \cite{wood07}. 
By including these parameters into the above equation, we obtain a rate coefficient of $2.44 \times  10^{-14}$ s$^{-1}$, which is 
roughly $4$ orders 
higher than the rate of external photo-dissociative reactions.  Cosmic ray induced photo-reactions can play an important role in 
 interstellar chemistry. The choices of these parameters are highly reaction specific and a wrong estimation 
might lead to misleading results.   In the UMIST 2006 database, $\gamma'$ ranges as high 
as $5290$ and as low as $25.0$. Though the higher values of $\gamma'$ are more reliable for the larger molecules because of the increasing overlap between the cross sections for photo-dissociation and the cosmic-ray-induced emission spectrum \citep{gred89},
 we use {\leavevmode\color{black} $\gamma'=752.0$} {\color{black} (in Table 1, we presented it under the column head marked $\gamma$)} 
for our calculations of the cosmic-ray-induced photo-dissociative reactions. 
The same photo-dissociation rate coefficients are adopted for the destruction of 
 OD-PA and its associated species.

Dissociative recombination (DR) reactions and rate coefficients for some of the associated ions of PA are shown in reactions R67-R75.
Pathways and rate coefficients of these reactions are adopted by
following the similar DR pathways available for $\mathrm{CH_3O}^+$, $\mathrm{CH_3OH_2}^+$, and
$\rm{CH_3OH^+}$ in \cite{wood07}. Since for all these reactions, $\beta \ne 0$, the reactions are temperature dependent.
The same DR rate coefficients are used for the associated ions of OD-PA.

{\leavevmode\color{black} 
For two or three-body gas-phase reactions in Table 1, the rate coefficients are represented in terms of the three
rate coefficients, $\alpha$, $\beta$ and $\gamma$. 
Most of the gas phase rate coefficients adopted here are either estimated or taken 
from similar kind of reactions. For reactions R1-R2, R4-R5 and R7-R8,
we assign $\beta$ and $\gamma$ to be zero. 
For the dissociative recombination reactions (R6 and R68-R75), these three coefficients are
estimated based on similar reactions listed  in \cite{wood07}. 
For reaction R3, 
we calculate the gas phase rate coefficient by using transition state theory, which leads to 
the Eyring equation \citep{eyri35}. Thus for reaction R3, these three parameters are not shown. 
For  destruction by photo-reactions, we supply these three coefficients following the
similar type of reactions available in \cite{wood07}.
For the destruction of polar neutral species by ions, we use the two relations discussed in 
\cite{su82}. These relations cannot be represented over the whole temperature range in terms of one set of three coefficients. 
However, we can tabulate $\alpha$, $\beta$ and $\gamma$ for reactions R9-R50 
in the low temperature regime.}

 \begin{figure}
 \hskip -1cm
 \centering{
 \vbox{
   \includegraphics[height=6cm,width=9cm]{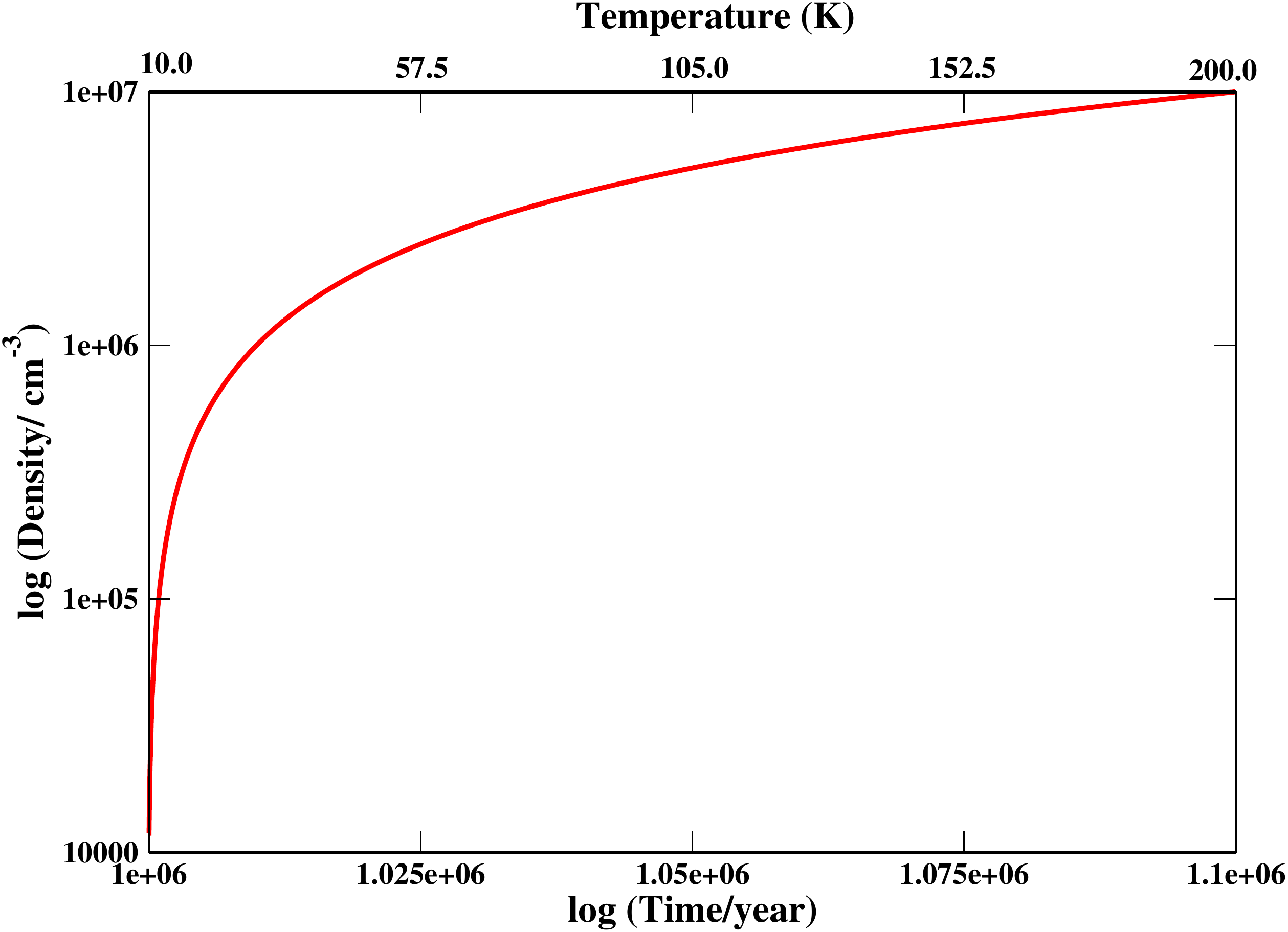}}}
 \caption{ {\leavevmode\color{black} Adopted physical conditions for the warm-up phase in which the density increases as the temperature increases.
.}}
 \end{figure}

\subsubsection{Ice phase rate coefficients}
Chemical enrichment of  interstellar grain mantles depends on the binding energies ($E_d$) 
and barriers against diffusion ($E_b$) of the adsorbed species. The binding energies of these 
species are often available from past studies \citep{alle77,tiel87,hase93,hase92}. But 
these binding energies mostly pertain to silicates. Binding energies of the 
most important surface species on ice, which mostly control the 
chemical composition of interstellar grain mantles,  are available from some 
recent studies \citep{cupp07,garr13}. We use these latter energies in our model. For the rest of 
the species for which binding energies are still unavailable, 
we use the same values as in past studies or estimate new values.  For barriers against diffusion, which are poorly known, we use values
equal to $0.35 E_d$  \citep{garr13}.  
 \begin{figure}
  \hskip -1cm
 \centering{
 \vbox{
   \includegraphics[height=9cm,width=9cm]{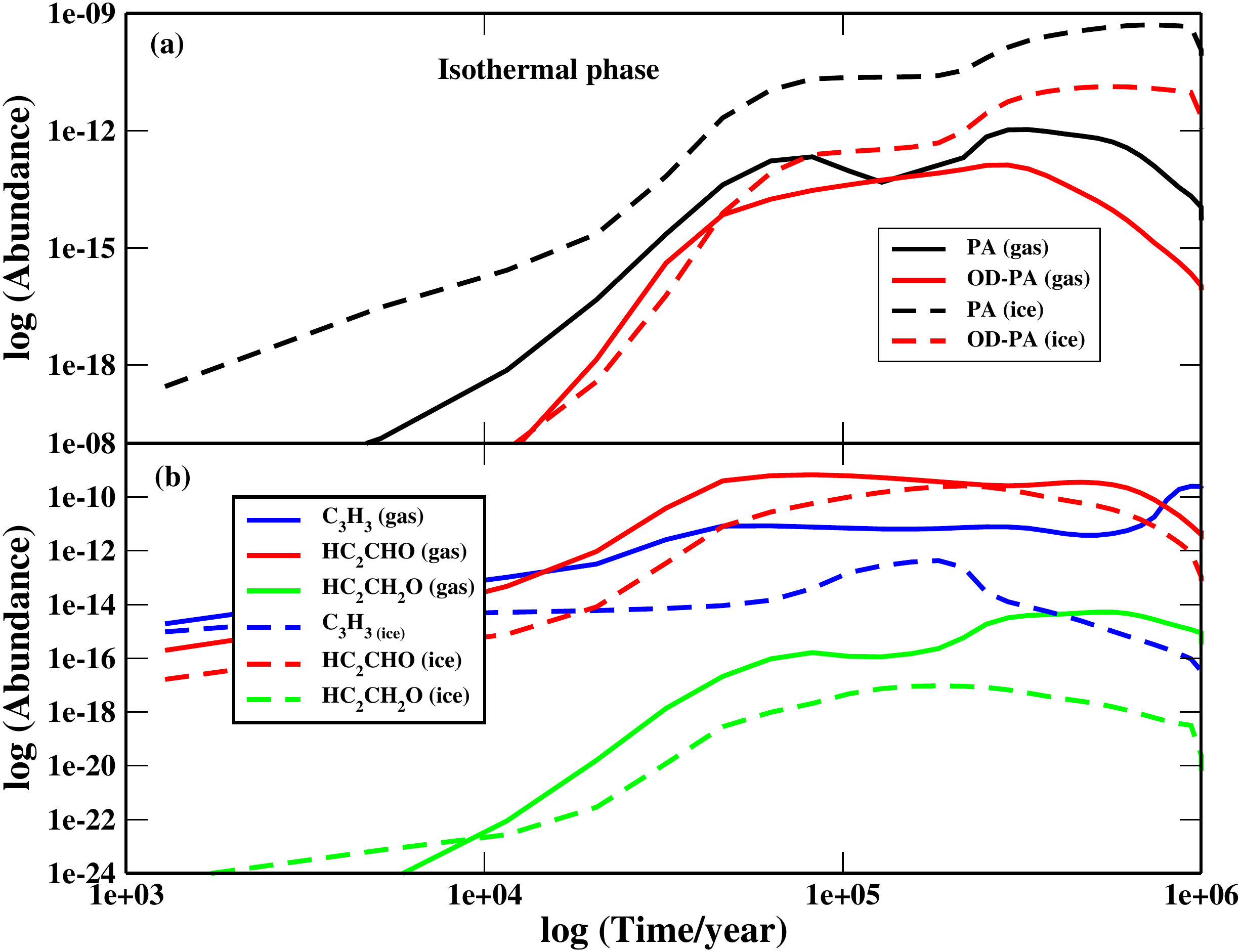}}}
  \caption{ \color{black} Chemical evolution of (a) PA and OD-PA and (b) their related species during 
the isothermal phase at $n_H = 10^4$ 
cm$^{-3}$ with a constant visual extinction of $10$.}
  \end{figure}

\begin{figure}
 \hskip -1cm
 \centering{
 \vbox{
  \includegraphics[height=9cm,width=9cm]{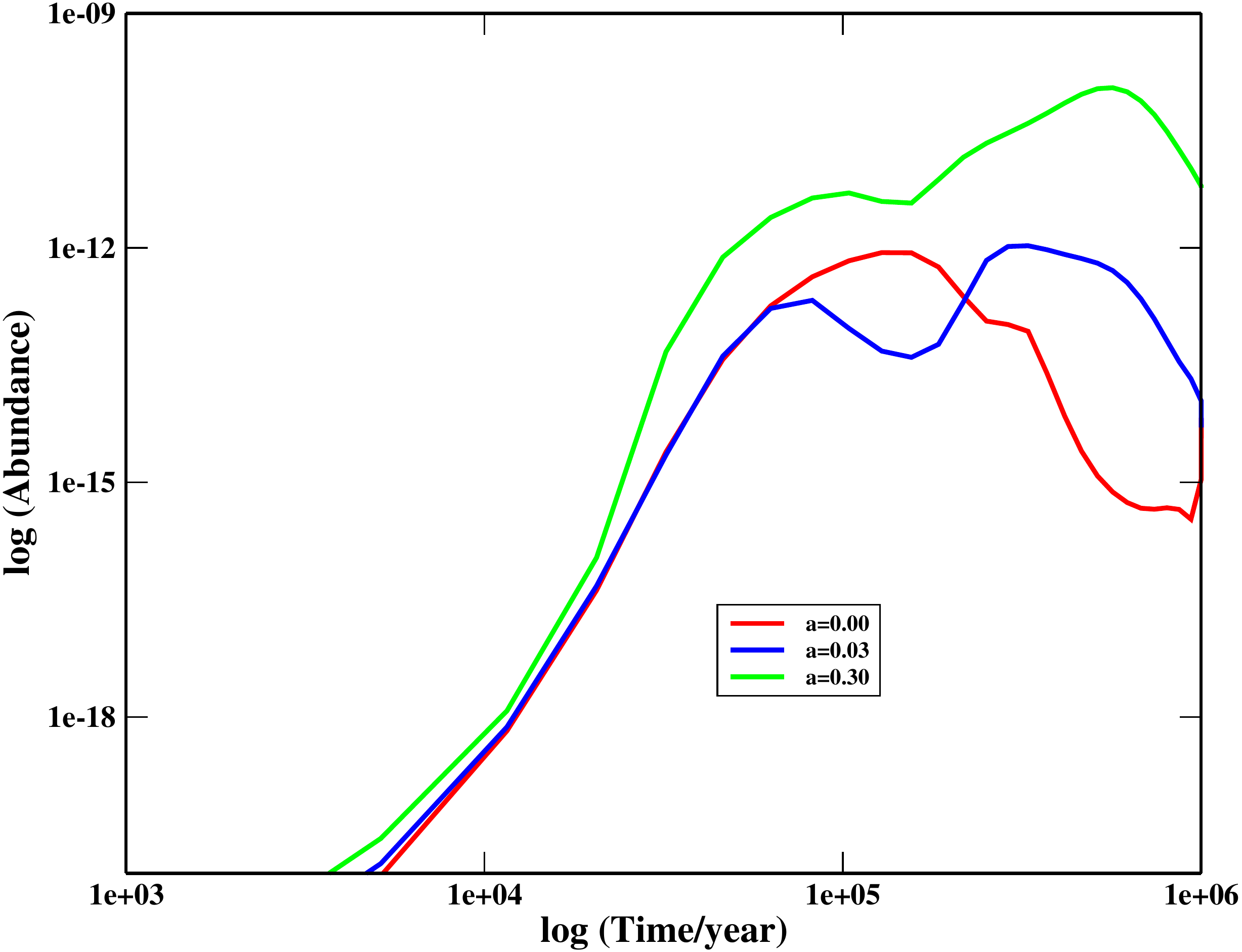}}}
 \caption{ Chemical evolution of PA during the constant density isothermal phase at $n_H = 10^4$ 
cm$^{-3}$ with a constant visual extinction of $10$ and
different values of the non-thermal reactive desorption parameter $a$.}
 \end{figure}

For the formation of PA in the ice phase, we include surface reactions R1-R5.
The rate coefficients ($R_{diff}$) of these reactions were calculated by the method
described in \cite{hase92}, which is based on thermal diffusion.  
They defined a probability factor $\kappa$ to differentiate between  exothermic reactions 
without activation energy barriers and reactions with activation energy barriers ($E_{\rm a}$) in such a way that
the effective rate coefficient becomes $R_{ij}=\kappa \times R_{diff}$.
The factor $\kappa$ is unity in the absence of a barrier.
For reactions with activation energy barriers, $\kappa$ is defined as the quantum mechanical
probability for tunneling  through a rectangular barrier of thickness a($=1$ $\AA$), which is calculated from the equation
\begin{equation}
\kappa= \exp [-2(a/ \hbar)(2\mu E_a)^{1/2}].
\end{equation}

{\leavevmode\color{black} 
Binding energies to the surface for some 
complex organics with a hydroxyl group are normally considered to be higher due to the phenomenon of hydrogen bonding \citep{coll04,latt11,garr13}.
Since PA has an -OH group, creating hydrogen bonds with a 
water substrate, this molecule is therefore expected to have a higher binding energy, 
close to that of water. Here, we assume the binding energy of both PA and OD-PA to be the same as methanol ($5530$ K). 
Since $\mathrm{HC_2CHO}$forms by the reaction between $\rm{C_3H_3}$ and $\rm{O}$,
we add the binding energies of $\mathrm{C_3H_3}$ ($2220$ K) and O ($800$ K) to determine
 a binding energy value $3020$ K. In the case of the binding energy of
$\mathrm{HC2CH2O}$, 
we add the binding energies of HC2CHO ($3020$ K) and H ($450$ K) to obtain
 $3470$ K.}
For other binding energies for species in reaction R1, R2, R3, R4 and R5,
we use $E_d(\mathrm{C_2H})= 1460$ K,
 $E_d(\mathrm{H_2CO})=2050$ K and $E_d(\mathrm{OH})=2850$ K. 
Our transition state theory  calculation found that reaction R3 (hydrogen addition to the carbon atom of the CHO group) 
contains an activation barrier 
of about $3.59$ kcal/mol ($1807$ K) in the ice phase.  
 
\begin{figure}
 \hskip -1cm
 \centering{
 \vbox{
  \includegraphics[height=10cm,width=9cm]{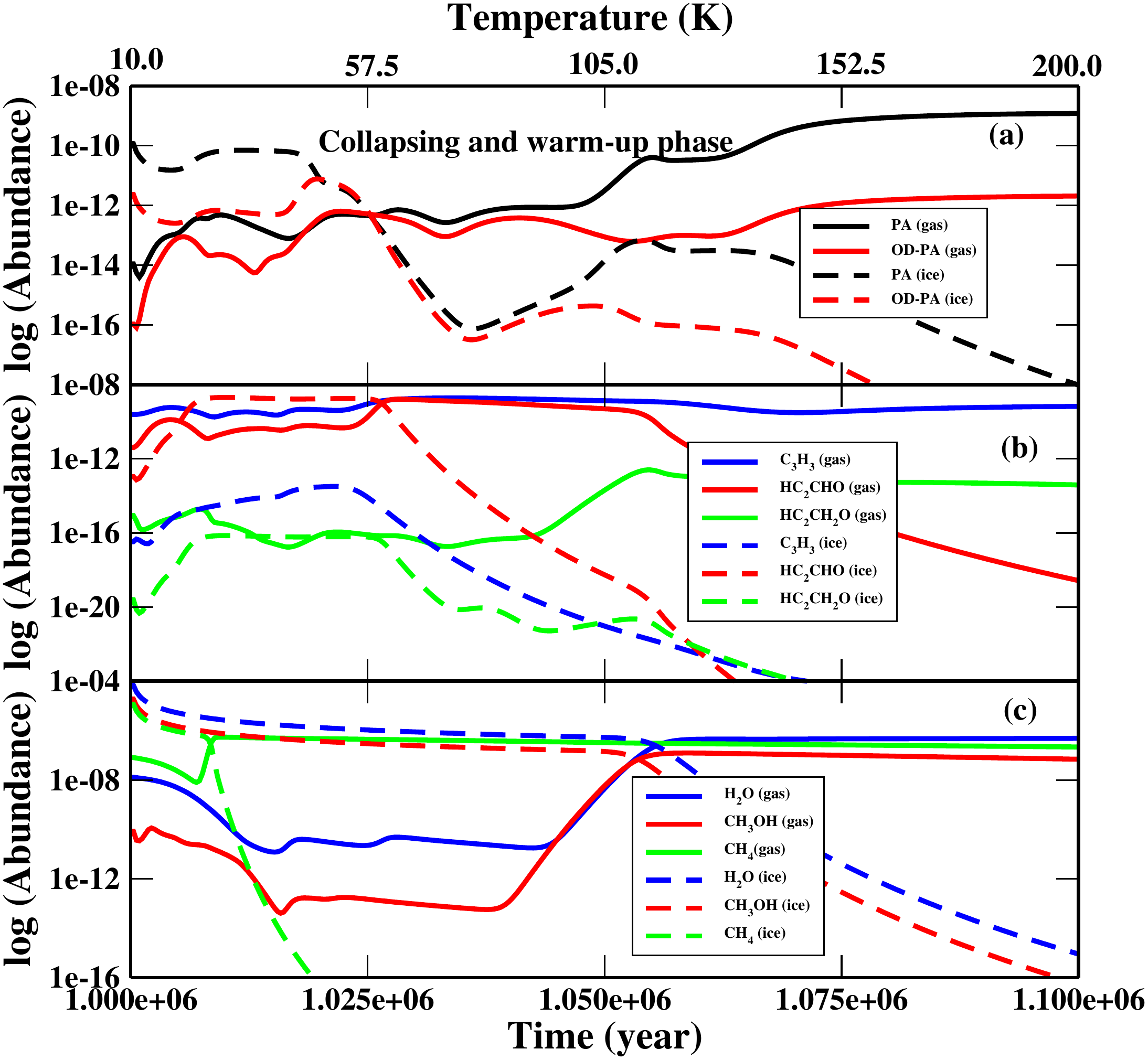}}}
 \caption{ {\color{black} Chemical evolution of (a) PA and OD-PA (b) $\mathrm{C_3H_3}$, $\mathrm{HC_2CHO}$ 
  and $\mathrm{HC_2CH_2O}$ {\color{black} (c) ${\rm H_2O}$, ${\rm CH_3OH}$ and ${\rm CH_4}$} for the warm-up phase, during which both density and temperature increase. }}
 \end{figure}

Destruction of ice phase PA occurs by various means; thermal desorption, 
non-thermal desorption from interstellar dust grains via exothermic surface reactions, 
cosmic ray induced desorption, photo-dissociation, and reaction with OH or OD radicals. 
Intact photo-desorption is not included. 
Thermal desorption plays a crucial role only at  high temperatures, depending upon the adsorbed species.
For thermal desorption, we use the relation prescribed by \cite{hase92}. 
At low temperatures, non-thermal desorption by reaction exothermicity, photons, and cosmic rays serve as 
important means for transferring the surface molecules to the gas phase. For 
non-thermal reactive desorption \citep{garr06}, we consider that all
surface reactions which result in a single product  break the surface-molecule bond with some
fraction $a$. Here, we use a fiducial value for the fraction of $a=0.03$. For the
cosmic ray induced desorption rates, we follow the expression developed by \cite{hase93}. 
Photo-dissociation by direct (R51-R58) as well as cosmic ray induced photons (R59-R66) are 
considered for the destruction of
ice phase PA and its two associated species, and the rate coefficients were assumed similar to 
their gas phase rate coefficients.  
 The dissociated products remain on the grain surface and are thus able to react 
and desorb depending on their binding energies.
For the destruction of ice phase PA by reactions with OH and OD  (R7 and R8), we consider the destruction pathways 
proposed by \cite{upad01} for the gas-phase, but we calculate the rate coefficients of these ice phase reaction pathways  
by using  diffusive reactions with barriers against diffusion, which have already been discussed.   
Selected ice phase rate coefficients used for PA are shown in Table 1 for $T=10$ K  and $100$ K.  Desorption and accretion rate coefficients are not included.   

\begin{table*}
\centering{
\scriptsize
\caption{Initial elemental abundances with respect to total hydrogen nuclei.}
\begin{tabular}{|c|c|}
\hline
Species&Abundance\\
\hline\hline
$\mathrm{H_2}$ &    $5.00 \times 10^{-01}$\\
$\mathrm{He}$  &    $1.40 \times 10^{-01}$\\
$\mathrm{N}$   &    $2.14 \times 10^{-05}$\\
$\mathrm{O}$   &    $1.76 \times 10^{-04}$\\
$\mathrm{e^-}$ &    $7.31 \times 10^{-05}$\\
$\mathrm{C^+}$ &    $7.30 \times 10^{-05}$\\
$\mathrm{S^+}$ &    $8.00 \times 10^{-08}$\\
$\mathrm{Si^+}$&    $8.00 \times 10^{-09}$\\
$\mathrm{Fe^+}$&    $3.00 \times 10^{-09}$\\
$\mathrm{Na^+}$&    $2.00 \times 10^{-09}$\\
$\mathrm{Mg^+}$&    $7.00 \times 10^{-09}$\\
$\mathrm{HD}$  &    $ 1.6 \times 10^{-05}$\\
\hline
\end{tabular}}
\end{table*}

\section{Chemical modeling}
\label{sec:chem}

\subsection{\leavevmode\color{black}Model}
We use our large gas-grain chemical network \citep{das15a,das15b} for the purpose of chemical modeling. 
We assume that the gas and grains are coupled through accretion and thermal and non-thermal
desorption processes. A visual extinction of $10$ and a cosmic ray ionization rate of 
$1.3\times 10^{-17}$ $\rm s^{-1}$ are used.
{\leavevmode\color{black}Including the deuterated reactions and deuterated species,} 
our present gas phase chemical network 
consists of $6384$ reactions involving $641$ gas phase 
species and the surface chemical network consists of $358$ reactions involving $291$ surface species.
The gas phase chemical network is mainly adopted from the UMIST 2006 database \citep{wood07}, but
 contains some deuterated reactions as well. To minimize the huge 	
computational time and avoid difficulty in handling a large chemical network, we have considered 
only some dominant deuterated reaction pathways. 
{\leavevmode\color{black}
Deuteration reactions important for the increase of deuteration in the cold gas phase are included
in our network.}
Our selection was based on some earlier 
studies by \cite{rodg96,robe00,robe03}. 
{\leavevmode\color{black} All the deuterated reactions, which were added/updated and identified
as the dominant pathways for the formation of essential deuterated species in the tables (Table 13-18) 
of \cite{albe13} are also included here.}
For the grain surface reaction network, 
we primarily follow \cite{hase92} and \cite{cupp07}, while for the ice phase deuterium fractionation reactions, 
we follow \cite{case02} and \cite{caza10}. In addition to this, we consider some reactions involving PA and coupled species, which are 
described in Section 2.  

Initial elemental abundances with respect to total H nuclei in all forms are shown in
Table 2. These ``low metal" abundances are often used for  dense clouds, where
hydrogen is mostly in the form of molecular hydrogen and ionization of the medium
is mainly governed by cosmic rays. We adopted these values from \cite{leun84}.
It is assumed that initially all deuterium is locked up in the form of HD. 
The initial abundance of HD  is assumed to be $1.6 \times 10^{-5}$ \citep{robe00} 
with respect to the total number of hydrogen nuclei.

{\leavevmode\color{black} In order to consider suitable physical conditions for a star forming region, 
we adopt a model with an initial phase of constant density ($\mathrm{n_H}$=$1.0\times 10^{4} \rm cm^{-3}$) 
and temperature ($T=10$ K) followed by a warm-up phase (temperature increases from $10$ K to 
$200$ K). 
The warm-up phase is assumed to be accompanied by an increase of density from 
$n_H = 10^4$ cm$^{-3}$ to $n_H = 10^7$ cm$^{-3}$, relevant when the material approaches 
the inner protostellar regions and appropriate for hot-core conditions.
Both phases have a visual extinction of $10$.
The initial phase is assumed to be sustained for $10^6$ years and 
the subsequent phase lasts for another $10^5$ years {\leavevmode\color{black}(a typical lifetime for a young embedded stage 
\citep{evan09})}.
In Fig. 1, the adopted time-dependent physical conditions for the second stage of evolution are shown.

For the rise in density and temperature, we assume constant slopes $m_{den}$ and $m_{temp}$ 
respectively for the density and temperature as determined from the following equations:
\begin{equation}
\rm m_{den}= \frac{\rho_{max}-\rho_{min}}{Time_f-Time_i}=99.9~{\rm cm^{-3} yr^{-1}}, 
\end{equation}
\begin{equation}
\rm m_{temp}= \frac{T_{max}-T_{min}}{Time_f-Time_i}=1.9 \times 10^{-3}~{\rm K yr^{-1}}. 
\end{equation}}


\subsection{\leavevmode\color{black}Modeling Results}
Figure 2a,b shows the chemical evolution of {\color{black} PA, OD-PA and their related species} in the gaseous 
(solid line) and ice phase (dashed line) during the cold isothermal phase. 
The ice phase production 
of PA is dominated by the successive H association reactions (R3 and R4). 
Similarly, for the case of  OD-PA, the association reactions are found to be dominant.
{\color{black} The peak abundances of PA and OD-PA and their related species are listed in Table 3, for both phases of our calculation. 
For example, in the isothermal cold phase, the peak fractional abundances of PA and OD-PA in the
ice p are \color{black} $5.03 \times 10^{-10}$
and \color{black} $1.33 \times 10^{-11}$ respectively and for the gas phase, these values 
are \color{black} $1.07 \times 10^{-12}$ and $1.32 \times 10^{-13}$  Note that the ice and gas phase peaks do not occur at the same time}.  

Non-thermal desorption processes significantly contribute to the maintenance of the 
gas-phase abundances
of PA and OD-PA with our fiducial value $a$ of $0.03$ for reactive desorption.  Fig. 3 show the gas 
phase abundance of PA versus time with $a=0$, $a=0.03$ and $a=0.3$ {\color{black} in the isothermal phase}.   
For the case of Fig. 3, throughout some but not all of the time of the evolutionary stage, 
the PA abundance for the $a=0.03$ case is slightly
higher than in the case without reactive desorption ($a=0$). 
For the highest `a' value ($a=0.3$), with the most rapid reactive desorption, the gas phase peak 
abundance is significantly larger than that with $a = 0.03$ and also with $a=0$.
{\color{black} In the isothermal phase, the peak abundance of gas phase PA for these three values of 
$a$ is found to be $8.69 \times 10^{-13}$, $1.07 \times 10^{-12}$ and $1.12 \times 10^{-10}$ 
for $a=0.00, \ 0.03$ and $0.30$ respectively.}
{\color{black} At the initial stages of evolution, the models with $a=0$ and $a=0.03$ give 
very similar gas phase abundances 
This means that the gas phase 
production of PA in our fiducial model 
is actually dominated by gas phase chemistry (by reaction R5) as long as the $a$ parameter is not 
too high.}

{\color{black} Figure 4a,b,c refers to the warm-up portion of our simulation. 
{\color{black} Fig. 4a represents the evolution of PA and OD-PA whereas Fig. 4b represents the 
evolution of PA related species such as ${\rm C_3H_3}$, $\rm{HC_2CHO}$ and
$\rm{HC_2CH_2O}$ and Fig. 4c represents the evolution of the major icy species water,
methane and methanol for the gas and ice phases.} 
{\color{black} The temporal evolution of gas phase and ice species shown in Fig. 4a,b,c 
appears to differ slightly with similar models found in \cite{garr08}.
This difference is solely due to the adopted physical condition mentioned in section 3.1
in comparison to that in \cite{garr08}.}
{\color{black} From Fig. 4c it is clear that among the main icy species, methane disappears 
first from the grain mantle due to its low adsorption energy ($1300$ K)
followed by methanol (adsorption energy $5530$ K) and water (adsorption energy $5700$ K).}

As can  be seen, the gas phase production of PA and OD-PA is found to be favourable around 
the high temperature region. This happens
due to the increase in the rate coefficient of reaction R3 with the temperature. 
For the ice phase reactions, H or D addition reactions become irrelevant beyond $40$ K, 
beyond which the $\rm{C_3H_3}$ and OH or OD radicals become mobile enough to react with 
each other on grain surfaces. {\color{black} At the same time, the destruction of PA 
and OD-PA by reactions with OH and OD increases due to the increase in the 
destruction rates with increasing temperature.}
Above $100$ K, $\rm{C_3H_3}$ and OH or OD radicals roughly disappear and  the only source of ice phase PA or OD-PA is accretion
from the gas phase. Since the density increases with time, accretion of the gas phase 
species becomes more favourable. 
Above $140$ K,  
the gas phase abundances of PA and OD-PA are roughly invariant. 

In Table 3, we list 
the peak abundances of PA, OD-PA and their related species for the warm-up phase along with their
corresponding times and temperatures. 
In this phase, the peak abundances of ice phase PA and OD-PA are found to be decreased in comparison 
with the isothermal phase. The peak fractional abundances 
of ice phase PA and OD-PA are found to be $7.09 \times 10^{-11}$ at
$32.2$ K,  and {\color{black} $7.23 \times 10^{-12}$ at $47.3$ K}, respectively, whereas,
for the gas phase, the corresponding peak abundances are $1.20 \times 10^{-9}$ 
and $2.14 \times 10^{-12}$ respectively.

Although reality is more complex
due to the consideration of variable density and temperature in the second phase,  
the basic explanation for the higher abundances of 
PA and OD-PA in the gas phase is that instead of chemical desorption, thermal desorption 
becomes much larger at higher temperatures, and indeed the ice abundances become 
infinitesimal at temperatures above $100$ K. 
In the first phase of our simulation (Fig. 2), which occurs while the 
temperature is $10$ K, only non-thermal slow desorption occurs, 
and indeed the ice abundances are higher.

For the gas-phase at higher temperatures, at which there is very little ice, fractionation occurs 
via gas-phase formation and destruction reactions, which lead to low fractionation at temperatures 
above $50$ K due to endothermic reactions between selected deuterated ions with H$_2$ which 
return deuterated species to normal ones.  After all, the high temperature limit of OD-PA/PA is 
$10^{-4}$, which is close to the D/H elemental value of $10^{-5}$.  
At temperatures under $50$ K, the abundance 
ratio of gaseous OD-PA/PA approaches unity, a very high degree of fractionation, while the ice ratio 
is $10^{-2}$.  
These high ratios probably stem from large values of the atomic abundance ratio D/H in the gas 
and on the grains at low temperatures.  
}

\begin{table*}
\scriptsize{
{\leavevmode\color{black}\caption{Peak abundances of PA, OD-PA and their related species with respect to H$_2$.}
\begin{tabular}{|p{0.40in}|p{0.50in}p{0.50in}|p{0.50in}p{0.50in}|p{0.50in}|p{0.30in}|p{0.50in}|p{0.50in}|p{0.30in}|p{0.50in}|}
\hline
{\bf Species}&\multicolumn{4}{|c|}{\bf Isothermal phase}&\multicolumn{6}{|c|}{\bf Warm-up phase}\\
\cline{2-11}
&\multicolumn{2}{|c|}{\bf gas phase}&\multicolumn{2}{|c|}{\bf ice phase}&\multicolumn{3}{|c|}{\bf gas phase}&\multicolumn{3}{|c|}{\bf ice phase}\\
\cline{2-11}
&{\bf Time (year)}&  {\bf Abundance}& {\bf Time (year)}& {\bf Abundance}& {\bf Time (year)}& {\bf T (K)} & {\bf Abundance}& {\bf Time (year)}& {\bf T (K)} & {\bf Abundance}\\
\hline
{\bf PA}&$3.29 \times 10^5$&$1.07 \times 10^{-12}$&$7.40 \times 10^5$&$5.03 \times 10^{-10}$&$1.10 \times 10^6$&200.0&$1.20 \times 10^{-09}$&$1.01 \times 10^{6}$&32.2&$7.09 \times 10^{-11}$\\
{\bf OD-PA}&$2.89 \times 10^5$&$1.32 \times 10^{-13}$&$5.66 \times 10^5$&$1.33 \times 10^{-11}$&$1.10 \times 10^0$&200.0&$2.14 \times 10^{-12}$&${\color{black}1.02 \times 10^{6}}$&${\color{black}47.3}$&${\color{black}7.23 \times 10^{-12}}$\\
{$\rm{\bf C_3H_3}$}&$9.36 \times 10^5$&$2.52 \times 10^{-10}$&$1.85 \times 10^5$&$4.32 \times 10^{-13}$&$1.034 \times 10^6$&75.0&$1.92 \times 10^{-9}$&$1.022 \times 10^{6}$&50.9&$3.25 \times 10^{-14}$\\
{$\rm{\bf HC_2CHO}$}&$8.22 \times 10^4$&$6.71 \times 10^{-10}$&$2.17 \times 10^5$&$2.60 \times 10^{-10}$&$1.028 \times 10^0$&63.5&$1.66 \times 10^{-9}$&$1.01 \times 10^{6}$&29.2&$2.05 \times 10^{-9}$\\
{$\rm{\bf HC_2CH_2O}$}&$5.14 \times 10^5$&$5.26 \times 10^{-15}$&$1.85 \times 10^5$&$9.43 \times 10^{-18}$&$1.054 \times 10^6$&114.1&$2.52 \times 10^{-13}$&$1.01 \times 10^{6}$&29.2&$7.23 \times 10^{-17}$\\
\hline
\end{tabular}}}
\end{table*}

In the warm-up case, the predicted fractional abundance of PA under hot-core conditions lies at approximately 
$10^{-9}$.
Our results argue strongly for the possibility of detection, especially at temperatures above 
{\color{black} 150 K}.  Interestingly, 
this prediction implies that PA does not occupy the same regions as does its isomer propenal.
(\cite{requ08} measured its abundance to $\sim 2.30 \times 10^{-9}$ with respect to H$_2$), but the species
 is seen in absorption against Sgr B2(N), 
and so is more likely to be present in colder foreground gas \citep{holl04}. For OD-PA, 
the predicted hot-core abundance is {\color{black}$\approx 10^{-12}$ at
temperatures above $150$.}  For the cold core 
case, even the peak fractional abundance of gas phase PA with the fiducial value of $a$ is sufficiently low to  rule out 
detection unless the highest value of the parameter
$a$ for reactive desorption is used.  We conclude that PA and possibly OD-PA can most 
likely be detected in warm-up regions at temperatures above 100 K. 

\section{\color{black} Astronomical Spectroscopy}
\label{sec:astro}
\subsection{\color{black} Detectability of PA and OD-PA in the millimeter-wave regime}
{\leavevmode\color{black} 
To estimate the possibility of detecting PA and OD-PA with present
astronomical facilities, we use the  CASSIS radiative transfer model [http://cassis.irap.omp.eu] at LTE with the JPL molecular database [http://spec.jpl.nasa.gov]. 
Propenal ($\mathrm{CH_2CHCHO}$) had already been successfully detected towards the high mass star forming region 
Sagittarius B2(N) with the $100$-meter Green Bank
Telescope operating in the frequency range $18-26$ GHz by \cite{holl04}. 
They had detected the $2_{11}-1_{10}$ line of $\mathrm{CH_2CHCHO}$ with $14$ mK intensity
($\approx$ 7$\sigma$ detection).  From our chemical model, we found that the peak abundances 
for PA and OD-PA are respectively $1.2\times10^{-9}$ and $
2.14\times10^{-12}$ with respect to $\rm{H_2}$ and for propenal, we use an abundance of
$2.3\times10^{-9}$ with respect to  $\mathrm{H_2}$ \citep{requ08}. 
As the input parameters, we use the following parameters, which
represent a typical high mass star forming region analogous to Sgr B2 (N): \\\\
\noindent Column density of $\mathrm{H_2}$= $10^{24}$cm$^{-2}$, \\
\noindent Excitation temperature ($T_{\rm ex}$)= $100$ K, \\
\noindent FWHM= $5$ km/s, \\
\noindent $\mathrm{V_{LSR}}$= $74$ km/s, \\
\noindent Source size to take into account the beam dilution= $3^{''}$

\begin{table*}
\centering{
\scriptsize
{\leavevmode\color{black}\caption{Calculated line parameters for mm-wave transitions of Propenal, PA and OD-PA with ALMA}
\begin{tabular}{|c|c|c|c|c|c|c|}
\hline
{\bf Species} & {\bf ALMA}& {\bf Frequency}&${\bf J_{k{_a}^{'}K_c^{'}}-J_{k{_a}^{''}K_c^{''}}}$&${\bf E_{up}}$&${\bf A_{ij}}$&{\bf Intensity}\\
&{\bf Band}&{\bf (GHz)}& & {\bf (K)}&&{\bf (mK)}\\
\hline
{\bf Propenal}&Band 1&44.4975&$5_{05}-4_{04}$&6.4&$4.33\times10^{-6}$&287\\
\hline
{\bf Propenal}&Band 2{\color{black}/Band 3}&89.05935&$10_{46}-9_{45}$&56.44&$3.06\times10^{5}$&2320\\
&&&$10_{47}-9_{46}$&&&\\
\hline
{\bf Propenal}&Band 3&115.80341&$13_{410}-12_{49}$&71.83&$7.34\times10^{-5}$&3920\\
&&&$13_{49}-12_{48}$&&&\\
\hline
{\bf PA}&Band 1&44.54410&$5_{05}-4_{04}$&6.42&$5.03\times10^{-7}$&8.18\\
\hline
{\bf PA}&Band 2{\color{black}/Band 3}&89.39184&$10_{65}-9_{64}$&72.01&$2.74\times10^{-6}$&102\\
&&&$10_{64}-9_{63}$&&&\\
&&89.39212&$10_{56}-9_{55}$&57.23&$3.21\times10^{-6}$&102\\
&&&$10_{55}-9_{54}$&&&\\
\hline
{\bf PA}&Band 3&116.23546&$13_{59}-12_{58}$&72.68&$8.10\times10^{-6}$&160\\
&&&$13_{58}-12_{57}$&&&\\
&&116.23586&$13_{86}-12_{85}$&125.04&$5.97\times10^{-6}$&160\\
&&&$13_{85}-12_{84}$&&&\\
\hline
{\bf OD-PA}&Band 1&44.88953&$5_{14}-4_{13}$&7.68&4.93E-7&0.013\\
\hline
{\bf OD-PA}&Band 2{\color{black}/Band 3}&87.65146&$10_{55}-9_{54}$&53.43&$3.43\times10^{-6}$&0.112\\
\hline
{\bf OD-PA}&Band 3&113.96386&$13_{59}-12_{58}$&68.58&$8.59\times10^{-6}$&0.206\\
&&113.96429&$13_{58}-12_{57}$&&&\\
\hline
\multicolumn{3}{c}{Band 1= 31-45 GHz}\\
\multicolumn{3}{c}{Band 2= 67-90 GHz}\\
\multicolumn{3}{c}{Band 3= 84-{\color{black}116} GHz}\\
\end{tabular}}}
\end{table*}
From our radiative transfer model, the intensity for $2_{11}-1_{10}$  line ($18.28065$ GHz) of PA 
is found to be $0.011$ mK,  which is far below than the 
detection limit of GBT (3$\sigma$$\approx$6mK). 
Because of the higher dipole moment components of propenal ($\mu_a=3.052$ D, 
$\mu_b = 0.630$ D and $\mu_c = 0$ D) compared with PA ($\mu_a=1.037$ D, $\mu_b = 0.147$ D 
and $\mu_c = 0.75$ D), intensity of PA is much lower compared with propenal. 
This prompted us to suggest the use of high-spatial and high-spectral resolution observations from the 
Atacama Large Millimeter/Sub-millimeter Array (ALMA). In particular, we can use lower frequency 
bands of ALMA -- Bands 1, 2 and 3 ($31-116$ GHz) --
since the observed rotational spectra for PA and OD-PA are expected to be cleaner in these frequency 
ranges given that low energy transitions of 
light molecules fall at much higher frequencies. This is why we have listed only the 
intense transitions for PA along with OD PA in Table 4  
that fall in ALMA Bands 1, 2 and 3.

{\color{black} Currently ALMA Band 3 is in operation while Bands 1 and 2 are not yet available.
The strongest transitions pointed out in Table 4 for PA (at $116.23546$ GHz and $116.23586$ GHz) 
are just beyond  ALMA Band 3 and are not available in the ALMA time estimator. Thus, we 
have focused on the next strongest transitions of Band 3 at 
$89.39184$ GHz and $89.39212$ GHz,  which also have an overlap with Band 2.
Using  Band 3, we can detect these transitions of PA with approximately 
$7.5$ hours of integration time by assuming a spectral resolution of $0.5$ km/s, 
a sensitivity of $17$ mK, a signal-to-noise ratio of $6$, an angular resolution $3$ arcseconds 
and $40$ antennas of $12$ meter array in the ALMA time estimator. 
But it is not possible to detect any lines for OD-PA using ALMA Band 3 with any reasonable integration time.}

\subsection{\color{black} Vibrational spectra}
Here we  use the  MP2 method with an aug-cc-pVTZ basis set for the computation of the vibrational 
transition frequencies and intensities of various normal modes of vibration of PA and OD-PA.
The prefix aug  stands for diffuse functions and cc-pVTZ refers to Peterson and Dunning’s correlation consistent  basis set \citep{pete02} for
performing our calculations. The influence of the solvent in vibrational spectroscopy is taken into 
account using the Polarizable
Continuum Model (PCM) with the integral equation formalism variant (IEFPCM) as a
default Self-consistent Reaction Field (SCRF) method  \citep{toma05, mier81}. The IEFPCM model is considered to be a
convenient one, because the second energy derivative is available for
this model and its analytic form is also available. Our calculation was performed in the presence of a 
solvent (water molecule) by placing the  PA or OD-PA solute in the cavity within the 
reaction field of the solvent. 
Since the dielectric constant of ice ($85.5$) is slightly higher than that of water ($78.5$), 
the vibrational frequencies reported here are not exactly those that pertain it ice but are close to these values.

In Table 5, we present the vibrational frequencies and intensities of PA and OD-PA in both the gas 
and ice phases. 
We show the band assignments and compare 
our results with the existing low resolution experimental results \citep{siva15, nyqu71}. 
Vibrational detections of molecules in the
interstellar medium may increase in importance with the impending launch of the the James  Webb  Space Telescope (JWST) although the spectroscopic
resolution may not be sufficient for rotational substructure in the gas. 

We find that the most intense mode of
PA in the gas phase appears  at $1070.33 \ \mathrm{cm^{-1}}$, which corresponds to CO stretching with an integral absorbance coefficient of  1.70$\times 10^{-17}$
cm $\mathrm{molecule^{-1}}$. 
This peak is shifted downward in the ice water phase by nearly $17$ cm$^{-1}$ and appears at $1053.89 \ \mathrm{cm^{-1}}$ with an
integral absorbance coefficient 2.46 $\times 10^{-17}$ cm $\mathrm{molecule^{-1}}$. Our frequencies can be compared with the experimental ice phase value of  $1030.7~\mathrm{cm^{-1}}$ \citep{siva15} and the vapor phase value of  $1051~\mathrm{cm^{-1}}$ \citep{nyqu71}. Other strong modes 
of vibrations are the OH torsion, CCC bending, $\mathrm{CH_2}$  wagging, CH stretching,  and OH stretching. 
In the case of OD-PA, the CCO stretching mode is the most intense mode and appears at $1068.32$ cm$^{-1}$ 
in the gas phase and $1050.98$ cm$^{-1}$ in the ice phase with integral absorbance coefficients of 
$1.61 \times 10^{-17}$ and $2.49 \times 10^{-17}$ cm molecule$^{-1}$ respectively.

Table 5 clearly shows the  differences between spectroscopic
parameters computed for propargyl alcohol between the two phases. These differences can be explained due to electrostatic effects, which are often
much less important for species placed in a solvent such as water  with high dielectric constant than they are in the gas phase but these changes will be 
significant if both the solvent and
the molecule are polar. 

The two sets of results for PA are in good agreement as regards frequencies 
with the existing  experimental results, as shown above for the single case of the strong CO stretching. 
Differences between the results could be attributed  to multiple reasons, one of which concerns the 
Gaussian 09 program, with which we are unable to consider the mixing of different modes under
harmonic oscillator approximations.  Another reason is that in our quantum 
chemical simulation of ice spectra, we considered a single propargyl alcohol molecule 
inside a spherical cavity and immersed it in a continuous medium with a dielectric constant, while in the experiment 
propargyl alcohol was deposited atop the ice, which could lead to the formation of  clusters.


\begin{table*}
\scriptsize{
\centering
\vbox{
\caption{Vibrational frequencies and intensities of PA and OD-PA in the gas phase and in water
ice at the MP2/aug-cc-pVTZ level of theory}
\begin{tabular}{|c|c|c|c|c|c|c|c|c|}
\hline
\hline
{\bf Species}&{\bf Peak }&{\bf Integral absorbance}&{\bf Peak }&{\bf Integral absorbance}&{\bf Band}&{\bf Experimental}\\
{}&{\bf positions }&\bf coefficient)&{\bf positions }& {\bf coefficient}&{\bf assignments}&{\bf values}\\

{}&{\bf (Gas phase)}&{($\mathrm{cm/molecule}$)}&{\bf (H$_2$O ice)}&{\bf( $\rm {cm/molecule}$})&&{\bf ($\mathrm{cm^{-1}}$)}\\
&\bf ($\mathrm{cm^{-1}}$)& & {\bf( $\mathrm{cm^{-1}}$})& &&\\
&&&& && \\
\hline
& 186.09 &1.48$\times 10^{-18}$ & 192.71 &1.09$\times 10^{-18}$    & CCC out of plane bending&$240^a$\\
& 257.59 &1.02$\times 10^{-17}$  &258.48 &1.59$\times 10^{-17}$  & OH torsion &\\
& 319.40 &9.89 $\times 10^{-18}$ & 330.49 &1.64$\times 10^{-17}$  & CCC bending&$305^a$\\
& 554.56 &1.91$\times 10^{-18}$ & 555.37& 2.97$\times 10^{-18}$ & CCO bending &$558.6^b$ $550^a$\\
& 640.33 &7.43 $\times 10^{-18}$ & 647.20  &1.08$\times 10^{-17}$ & CH bending & $639^a$\\
& 665.86 & 6.09$\times 10^{-18}$ & 672.76  &8.99$\times 10^{-18}$ & CH out of plane bending &$654^b$ $657^a$\\
& 922.65 & 4.37$\times 10^{-18}$ & 922.89  & 5.97$\times 10^{-18}$ &C-C stretching&$916.8^b$ $918^a$\\
& 1002.52 & 2.89$\times 10^{-18}$ & 987.69 & 6.28$\times 10^{-18}$ & CH$_2$ rocking&$987.2^b$ $980^a$\\
& 1070.33& 1.70$\times 10^{-17}$ & 1053.89   & 2.46$\times 10^{-17}$ & C-O stretching&$1030.7^b$ $1051^a$\\
{\bf PA}& 1225.86 & 2.61$\times 10^{-18}$& 1224.44  &3.85$\times 10^{-18}$& CH$_2$ twisting &$1228^a$\\
& 1355.10 & 4.50$\times 10^{-19}$ & 1371.30  & 9.84$\times 10^{-19}$ & CH/OH bending&---\\
& 1417.76& 8.47$\times 10^{-18}$& 1409.79 & 1.03$\times 10^{-17}$ & CH$_2$ wagging &$1362^b$\\
& 1517.33 & 2.70 $\times 10^{-19}$& 1505.02   & 5.81$\times 10^{-19}$ & CH$_2$ scissoring&$1480^a$\\
& 2131.06 &2.92$\times 10^{-19}$ &  2125.63 & 4.23$\times 10^{-19}$ & CC stretching& $2185^a$\\
& 3066.90 & 4.64$\times 10^{-18}$ & 3084.99  &4.92$\times 10^{-18}$ & CH$_2$ symmetric stretching&$2922^b$ $2930^a$\\
& 3145.17 & 8.51$\times 10^{-19}$& 3154.35 & 1.05$\times 10^{-18}$ & CH$_2$ antisymmetric stretching&$2940^a$\\
& 3483.14 & 9.39$\times 10^{-18}$ & 3364.13 &1.45$\times 10^{-17}$  & C-H stretching&$3282^b$ $3322^a$\\
& 3831.50 & 6.97 $\times 10^{-18}$& 3813.81  &1.32$\times 10^{-17}$ & O-H stretching&--- \\
\hline
&164.70  &3.97$\times 10^{-18}$ & 173.05 &5.27$\times 10^{-18}$    & OD torsion&\\
& 221.58 &5.59$\times 10^{-18}$  &221.28 &9.19$\times 10^{-18}$  & CCC out of plane bending &\\
& 307.17 &3.08 $\times 10^{-18}$ & 313.58 &5.01$\times 10^{-18}$  & CCC bending&\\
& 543.25 &2.13$\times 10^{-18}$ & 547.08& 3.24$\times 10^{-18}$ & CCO bending &\\
& 640.32 &7.35 $\times 10^{-18}$ & 647.20 &1.08$\times 10^{-17}$ & CH bending &\\
& 665.39& 6.12$\times 10^{-18}$ & 672.72 &9.03$\times 10^{-18}$ & CH out of plane bending &\\
& 885.79 &5.89$\times 10^{-18}$ & 852.99 & 8.03$\times 10^{-18}$ &OH bending&\\
& 961.44 &2.40$\times 10^{-18}$ & 944.47 & 3.55$\times 10^{-18}$ & CC stretching &\\
& 1068.32&1.61$\times 10^{-17}$ & 1050.98 & 2.49$\times 10^{-17}$ & CCO stretching&\\
{\bf OD-PA}& 1086.19 & 5.05$\times 10^{-19}$& 1086.55 &1.06$\times 10^{-18}$& CH$_2$ rocking&\\
& 1303.40 & 2.16$\times 10^{-20}$ & 1313.89 & 8.66$\times 10^{-20}$ & CH$_2$ twisting&\\
& 1396.85& 4.39$\times 10^{-18}$& 1396.93 & 6.64$\times 10^{-18}$ & CH$_2$ wagging & \\
& 1517.29 & 2.63 $\times 10^{-19}$&1504.99 & 5.71$\times 10^{-19}$ & CH$_2$ scissoring&\\
& 2131.04 &2.94$\times 10^{-19}$ &  2125.63 & 4.25$\times 10^{-19}$ & CC stretching&\\
& 2789.51 & 4.50$\times 10^{-18}$ & 2776.81 &8.25$\times 10^{-18}$ & OD stretching&\\
& 3066.98 & 4.54$\times 10^{-18}$& 3085.06& 4.83$\times 10^{-18}$ & CH$_2$ symmetric stretching &\\
& 3145.44 & 7.29$\times 10^{-19}$ & 3154.61 &8.76$\times 10^{-19}$  &CH$_2$ antisymmetric stretching &\\
& 3483.14& 9.40 $\times 10^{-18}$& 3464.13 &1.46$\times 10^{-17}$ & CH stretching& \\
\hline
\multicolumn{4}{c}{$^a$ \cite{nyqu71}} (vapor phase), $^b$ \cite{siva15} (ice phase experiment)\\
\end{tabular}}}
\end{table*}

\section{Conclusions}
\label{sec:concl}

Earlier results suggest that PA could be treated as a precursor of benzene formation \citep{wils03}.
The interstellar identification of propenal, which is an isomer of PA, encouraged us to check 
the possibility of detecting PA and OD-PA in the ISM.   Our quantum chemical calculations, combined with existing rotational transition frequencies for PA and OD-PA, allow us to achieve the following major results for this paper:  

\noindent {$\bullet$ A complete reaction network has been prepared for the formation and destruction of
PA. Various formation routes are discussed and identified based on calculated exothermicities and endothermicities. 
}

\noindent {$\bullet$ The predicted abundances of PA and OD-PA  in the gas and ice phases yield some  
information for both cold and warm interstellar sources on the possibility of observing this molecule in the ISM, especially in hot-core regions such as Sgr B2(N). }

\noindent{$\bullet$ \color{black} A simple radiative transfer model has been employed to discuss the 
possibility of detecting PA in the ISM through millimeter-wave observations. The most likely environments for detection of PA are hot-core regions with a temperature above 100 K.
}

\noindent {$\bullet$  The frequencies and intensities of vibrational transitions of PA and OD-PA 
both in the gas and in a mimetic for an ice environment are
calculated and compared with existing low resolution  experimental frequencies.  The calculated intensities should be particularly useful.}

\section{Acknowledgments}
PG, AD \& SKC are grateful to DST (Grant No. SB/S2/HEP-021/2013) and ISRO Respond 
(Grant No. ISRO/RES/2/402/16-17) for financial support. 
LM thanks the ERC for a starting grant (3DICE, grant agreement 336474). E. H. wishes to 
acknowledge the support of the National Science Foundation for his astrochemistry program.

\end{document}